\documentclass[aps,preprint,onecolumn,floatfix,superscriptaddress,nofootinbib]{revtex4-1}
\usepackage{amsmath}
\usepackage{graphicx,bm}
\usepackage[colorlinks,citecolor=blue]{hyperref}
\usepackage{cancel}
\usepackage{ulem}

\begin{document}
\title{Constraining light dark matter and mediator with $B^+ \rightarrow K^+ \nu \bar \nu$ data}

\author{Murat Abdughani}
\email{mulati@xju.edu.cn} 
\author{Yakefu Reyimuaji}
\email{yreyi@hotmail.com}
\affiliation{School of Physical Science and Technology, Xinjiang University,Urumqi 830017, China}
\begin{abstract}
We study the decay of $B^+$ meson into $K^+$ plus a light mediator $\phi$, which subsequently decays into a dark matter pair, $\bar \chi \chi$. Integrating constraints from DM relic density, direct detection, collider data and $B$-physics, alongside the recently reported results form Belle II experiment, we analyze the couplings between the mediator, standard model fermions, and the dark matter particles. Our results indicate that if the decay process $\phi \rightarrow \bar \chi \chi$ is kinematically allowed, i.e. $m_\phi > 2m_\chi$, then the mediator mass must be constrained within 0.35 GeV $\lesssim m_\phi \lesssim$ 3 GeV. Conversely, if $m_\phi < 2m_\chi$, the mediator $m_\phi$ is long-lived relative to the detector size, and the only allowed decay channel is $\phi \rightarrow e^+ e^-$. 
\end{abstract}
 
\date{\today}
\maketitle
\newpage

\section{Introduction} 

Rare $B$-meson decay with a kaon and neutrino pair in the final state, i.e. $B^+ \rightarrow K^+ \nu \bar \nu$, has a clear theoretical prediction among the flavor changing neutral current processes. It is severely suppressed due to the loop effect and the Glashow-Iliopoulos-Maiani mechanism \cite{Glashow:1970gm}. Thus it plays an important role in searching for new physics and testing the Standard Model (SM). While the SM prediction for the the branching fraction is $(5.58 \pm 0.37) \times 10^{-6}$ \cite{Parrott:2022zte}, the Belle II experiment at the SuperKEKB asymmetric energy electron-positron collider has recently reported the value of $(2.3 \pm 0.7) \times 10^{-5}$ \cite{Belle-II:2023esi} with the significance of about $3\sigma$ above the SM prediction. This discrepancy may call for the new physics beyond the SM (BSM).

Although the weak interaction of neutrinos makes it invisible at the detector, processes with missing energy $\cancel E$ in the final state, i.e. $B^+ \rightarrow K^+ + \cancel E$, may not contribute to measured branching fraction of the decay $B^+ \rightarrow K^+ \nu \bar \nu$  in Ref. \cite{Belle-II:2023esi}. This decay process can be reinterpreted to the two-body decay process $B^+ \rightarrow K^+ X$ \cite{Altmannshofer:2023hkn}, where $X$ is stable or decay invisibly. In this interpretation, they achieved maximum branching ratio of $Br[B^+ \rightarrow K^+ X] = (8.8\pm2.5)\times 10^{-6}$ with a significance of 3.8$\sigma$ at $m_X \simeq 2$ GeV. Considering the no excess in BaBar experiment, the combined result is $Br[B^+ \rightarrow K^+ X] = (5.1\pm2.1)\times 10^{-6}$ with a significance of 2.4$\sigma$. Similar two-body decay processes have also been studied at Refs. \cite{MartinCamalich:2020dfe,Ferber:2022rsf}. Dark matter (DM) or any other light particles with sufficiently weak interaction strength involved in this process can be considered as the missing energy. Therefore, precision experiments of $B$ meson decay are crucial in the searching or constraining the DM and other light BSM particles. 

DM on the other side, is an indispensable ingredient for the evolution of the Universe. The thermal DM known to be in equilibrium with SM particles in the early times and freeze-out later when its annihilation rate could not catch up with the expansion rate of the Universe \cite{Srednicki:1988ce}. After the thermal freeze-out, the DM comoving number density remains constant and leads to the observed DM relic abundance. This, on the other hand, determines the DM annihilation rate at the freeze-out. Interestingly, annihilation cross section of this process is of the same order of magnitude as the weak interaction. Thus, this weakly interacting massive particle (WIMP) \cite{Ellis:1983ew} has become the mostly well-studied and promising candidate for DM. The lower limit for the mass of the WIMP, which is model dependent, is conventionally a few GeV \cite{Lee:1977ua}. However, the null result from various DM search experiments have pushed the lower limit to MeV scale with proper parameter or model selection \cite{Pospelov:2007mp}.

Combined study of DM and new physics hint from $B$-meson decay is an intriguing window towards the construction of BSM physics. In this work, an extension of the SM with a light scalar mediator and a Majorana fermion is studied. By integrating constraints from various experiments, we delineate the optimal parameter ranges for our model. The paper is structured as follows. In Sec.~\ref{sec:DMmodel} a simple DM model is presented. Section~\ref{sec:Bdecay} gives discussions about the $B$-meson decay into kaon and SM fermions as well as extra fermionic DM pairs. Experimental constraints and parameter spaces are studied in Sec.~\ref{sec:DMrlcden}, followed by results and discussions in Sec.~\ref{sec:results}. Finally, the conclusion of the paper is drawn in Sec.~\ref{sec:conclud}.

\section{A simple DM model} \label{sec:DMmodel}

\begin{figure}
    \centering
    \includegraphics[width=0.3\textwidth]{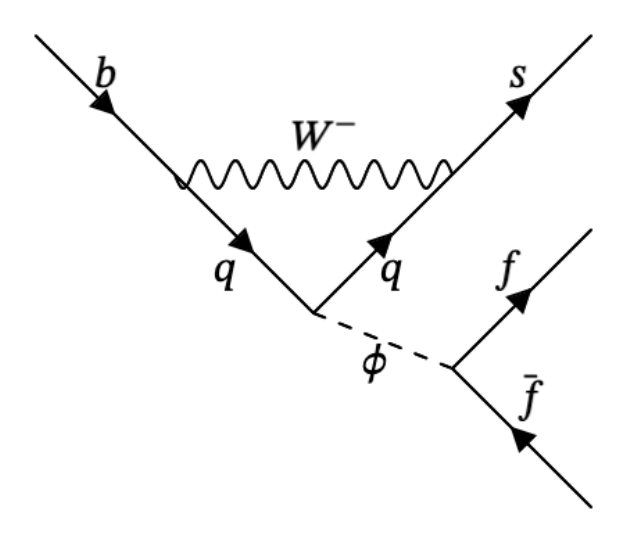}
    \includegraphics[width=0.3\textwidth]{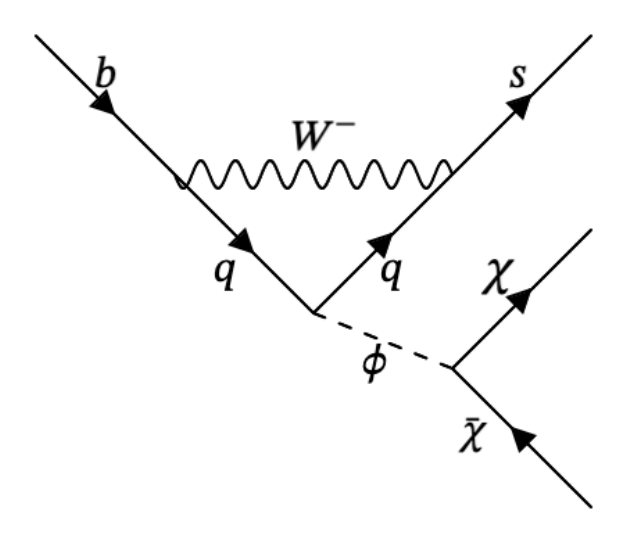}
    \caption{The BSM leading-order Feynman diagrams for $b \to s \chi \bar \chi$ and $b \to s f \bar f$ processes. The $q$ stands for $u,c,t$ quarks, $f$ include all quarks and massive leptons.}
    \label{fig:feynman}
\end{figure}

In this section, we study a generic model featuring light Majorana DM $\chi$ and a light scalar $\phi$ that couples with the SM fermions. The dimension-4 interaction Lagrangian can be simply written as
\footnote{Pseudoscalar current $\bar f \gamma^5 f$ vanishes at the loop order for $0^- \rightarrow 0^-$ meson decay processes \cite{He:2022ljo,Li:2021sqe}.}
\begin{equation}
    \mathcal{L_{\rm int}} = - \frac{y m_f}{v} \phi \bar f f - \frac{1}{2} \kappa \phi \bar \chi \chi \ ,
    \label{eq:lagrangian}
\end{equation}
where $m_f$ represent the masses of the SM fermions, $y$ is the weight of the Yukawa coupling, $v \simeq 246$ GeV is the electroweak vacuum expectation value. The Feynman diagrams at the leading-order for DM and fermion pair productions in the final state are illustrated in FIG. \ref{fig:feynman}.

We note that the simple Lagrangian in Eq.\eqref{eq:lagrangian} deos not originate from gauge symmetry but is phenomenologically viable after the electroweak symmetry breaking of higher-dimensional effective operators \cite{Kamenik:2011vy}. The constant $\frac{m_f}{v}$ in the coupling indicates that the this operator is induced from integrating out the Higgs portal \cite{Patt:2006fw}. Effective field theory approaches to the decay processes of $B$ meson with a single invisible scalar or fermion pair in the final state in dimension-5 or dimension-6 operators have been studied in Refs. \cite{Li:2021sqe,1809.01876,Krnjaic:2015mbs,Filimonova:2019tuy,Kachanovich:2020yhi,He:2022ljo}. The anomaly is also explained by the UV complete models~\cite{Ovchynnikov:2023von,Asadi:2023ucx,Athron:2023hmz}.

\section{$B$-meson decay}\label{sec:Bdecay}

When the scalar mediator mass, $m_\phi$, is smaller than the mass difference between the $B$ and $K$ meson, i.e. $m_\phi < m_B - m_K$, the effective Lagrangian for the $b \rightarrow s \phi$ process can be obtained by integrating out heavy particles running in the loop shown in the FIG. \ref{fig:feynman}. This process is described by the following Lagrangian \cite{Schmidt-Hoberg:2013hba, Batell:2009jf}

\begin{equation}
    \mathcal{L}_{b \rightarrow s\phi} = \frac{y m_b}{v} \frac{3 \sqrt{2} G_F m_q^2 V^*_{qs} V_{qb}}{16 \pi^2} \times \phi \bar s_L b_R  + {\rm h.c.} \ ,
    \label{eq:effLag}
\end{equation}
where $q$ denotes the $u,c,t$ quarks in the loop, $G_F$ is the Fermi constant, and $V_{qs,qb}$ are the Cabibbo-Kobayashi-Maskawa (CKM) matrix elements. It is obvious that contribution from the quarks other than the top are negligible.

The decay width for the process $B \rightarrow K +\phi$ is given by
\begin{align}
    \Gamma_{B\rightarrow K \phi} =& \left(\frac{y m_b}{v} \frac{3 \sqrt{2} G_F m_t^2 V^*_{ts} V_{tb}}{16 \pi^2} \right)^2 \frac{\sqrt{(m_B^2-(m_K+m_\phi)^2)(m_B^2-(m_K-m_\phi)^2)}}{16\pi m_B^3} \nonumber \\
    \times & |\langle K|\bar s_L b_R|B \rangle|^2 \ ,
    \label{eq:width}
\end{align}
where the hadronic transition matrix element is
\begin{equation}
    \langle K|\bar s_L b_R|B \rangle = \frac{m_B^2-m_K^2}{2(m_b-m_s)}f_0(q^2) \ , 
\end{equation}
and $m_{b,s}$ are the masses of the bottom and strange quark, respectively. The $f_0(q^2)$ is the form factor, with $q^2 = m_\phi^2$ for the final-state particle $\phi$. We adopted two different form factor schemes from Ref. \cite{2301.06990} and Ref. 
\cite{2304.12837} for the stability of our result, and details of which are discussed in the Appendix \ref{appendix:FF}. Our analysis reveals that the results obtained from two schemes are merely distinguishable; hence, we present only the results from the first scheme in the main text.

Subsequently, if kinematically allowed, the scalar $\phi$ may decay into DM or SM fermion pairs. The decay width of $\phi$ to a fermion pair, including DM, is
\begin{equation}
    \Gamma_{\phi \rightarrow \bar f f} = \frac{C_{\phi \bar f f}^2}{8\pi}  m_\phi \left(1-\frac{4m_f^2}{m_\phi^2} \right)^{3/2} \Theta (m_\phi - 2m_f) \ ,
\end{equation}
where $C_{\phi \bar f f}^2$ is scalar-fermion-fermion interaction coefficient, and $\Theta$ is the Heaviside step function. For the decay width to the hadronic final states, we consulted to the Refs. 
\cite{Schmidt-Hoberg:2013hba, McKeen:2008gd, 1809.01876}.

\begin{figure}
    \centering
    \includegraphics[width=0.47\textwidth]{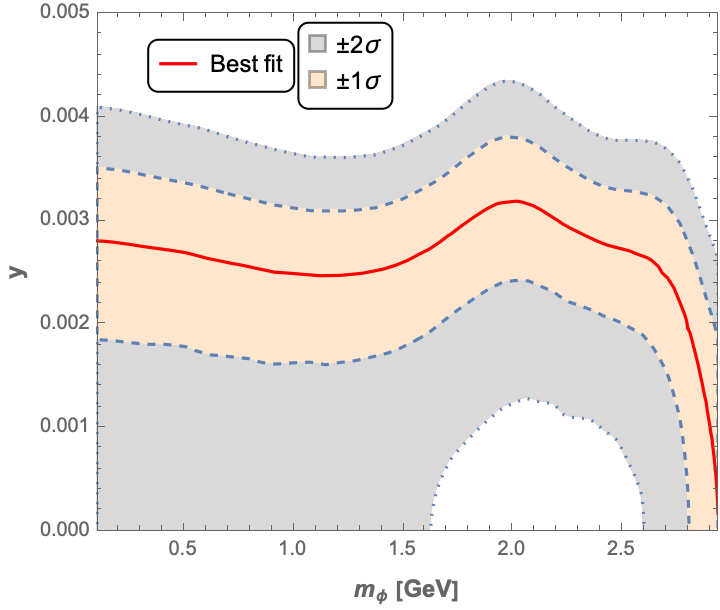}
    \hspace{0.5cm}
    \includegraphics[width=0.45\textwidth]{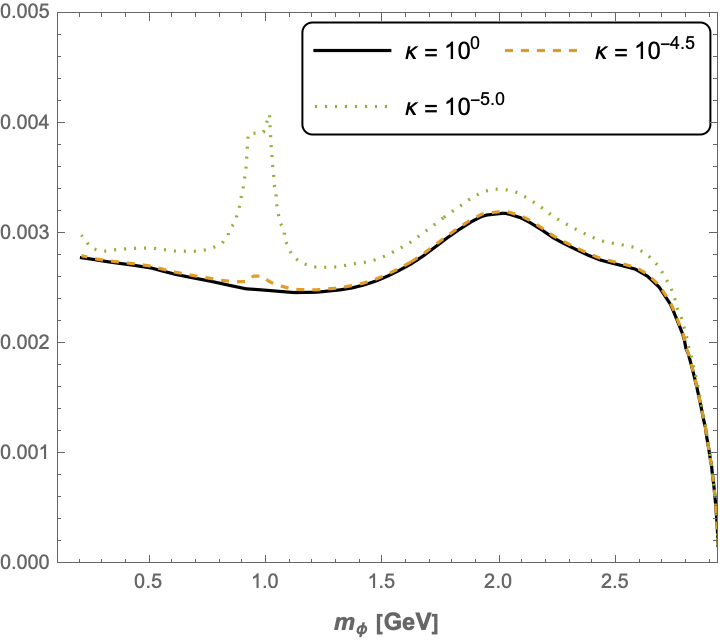}
    \caption{Left: Best-fit line (red solid line), with 1$\sigma$ (light yellow) and 2$\sigma$ (light gray) shaded areas, satisfying the branching fraction of $B \rightarrow K + \phi$ as shown in FIG. 1 of Ref. \cite{Altmannshofer:2023hkn} in $(m_\phi, y)$ space. Right: The branching fractions of process $B\rightarrow K + \phi$ followed by $ \phi \rightarrow \bar \chi \chi$ align with the best-fit line in the left panel for different $\kappa$ values, where DM masses are fixed at 0.1 GeV.}
    \label{fig:br}
\end{figure}

In the left panel of FIG. \ref{fig:br}, we show the coupling $y$ as a function of $m_\phi$, adopting the branching fraction of $B \rightarrow K + \phi$ from FIG. 1 of Ref. \cite{Altmannshofer:2023hkn}, assuming that branching fraction of $\phi$ decays into $\chi \bar \chi$ is 100\%. In the right panel, with $m_\chi$ fixed at 0.1 GeV, we draw curves for different $\kappa$ values that achieve the same branching ratios as the red solid line in the left panel. For the fixed $m_\chi$ and $m_\phi$, as the value of $\kappa$ decreases, the branching fraction $Br[\phi \rightarrow \bar \chi \chi]$ also decreases. Consequently, a larger $y$ is required to increase $Br[B\rightarrow K + \phi]$ so that the overall $Br[B\rightarrow K+\bar \chi \chi] \equiv Br[B\rightarrow K + \phi] \times Br[\phi \rightarrow \chi \bar \chi]$ aligns with the desired values. The bump near $m_\phi \sim 1$ GeV is due to the peak in the decay width to SM particles shown as in FIG. 4 of Ref. \cite{1809.01876}, which is caused by a pion pair in the final state.

\section{DM Relic Density and Constraints}\label{sec:DMrlcden}

When the decay process $\phi \rightarrow \bar\chi \chi$ is kinematically allowed, that is, $m_\chi < m_\phi/2$, in the vicinity of freeze-out, DM can only annihilate into lighter SM fermion pairs through an $s$-channel heavy mediator $\phi$. The cross section which determines the DM relic density for the process $\bar \chi \chi \rightarrow \bar f f$ is given by
\begin{equation}
   \sigma_{\bar \chi \chi \rightarrow \bar f f} = \frac{1}{512 \pi} \left( \frac{\kappa y m_f}{v} \right)^2 \sqrt{1-\frac{4m_f^2}{s}} \left(1-\frac{2m_f^2}{s} \right) \left(1-\frac{2m_\chi^2}{s} \right) \frac{s}{(s-m_\phi^2)^2} \, ,
\end{equation}
where $s \equiv E^2_{\rm cm}$ is the center-of-mass energy squared. Note that this equation holds for nonresonant annihilation. The thermally averaged cross section is defined as~\cite{Gondolo:1990dk}
\begin{equation}
    \langle \sigma v \rangle_{\bar \chi \chi \rightarrow \bar f f} = \int_{4m^2_\chi}^\infty ds \ \sigma_{\bar \chi \chi \rightarrow \bar f f} v_{\rm lab} \frac{\sqrt{s-4m_\chi^2}(s-2m_\chi^2)K_1(\sqrt{s}/T)}{8m_\chi^4 T K_2^2(m_\chi/T)} \ ,
\end{equation}
where $K_i$ are the modified Bessel functions of order $i$. If DM does not resonantly annihilate, the possible maximum of $\langle \sigma v \rangle$ is at least three orders less than the standard thermal cross section of $10^{-26}$ cm$^3/s$. Therefore, we have to resort to resonantly enhanced cross section of DM annihilation. We utilize \texttt{Feynrules} \cite{Alloul:2013bka} to implement the models, and then import them to \texttt{MicroMEGAS-5.3.41} \cite{Belanger:2010gh} for the calculation of DM relic density and DM-nucleus cross section.

\begin{table}[t]
    \centering
    \begin{tabular}{cccc}
\hline\hline
Parameters&Minimum&Maximum & Prior \\
\hline
$m_\phi$&0.01 GeV\ \ &$m_B-m_K$\ \ &flat/log\\
$m_\chi$ & 0.01 GeV& $m_\phi/2$& flat/log \\
$\kappa$ & $10^{-6}$ & 4 & log \\
y & $10^{-6}$ & 0.1 & log \\
\hline \hline
\end{tabular}
\caption{Ranges and priors for input parameters adopted in the scans.}
\label{tab:range}
\end{table}

To explore full parametric phase space, we scan over the range specified in TABLE \ref{tab:range},
where upper limit of 0.1 for $y$ is taken from LEP \cite{L3:1996ome}. We evaluate the likelihoods associated with DM relic density measurements from PLANCK and direct detection (DD) results from various experiments. The $\chi^2_{\rm scan}$ is defined as the sum of the individual $\chi^2$ values from the DM relic density and the DD processes: 
\begin{equation}
    \chi^2_{\rm scan} = \chi^2_{\Omega h^2} + \chi^2_{\rm DD}\ .
\end{equation}
We hire emcee \cite{Foreman-Mackey:2012any} based on Markov Chain Monte Carlo (MCMC) method to undertake the task of sampling the parameter space with the likelihood $ \propto {\rm exp}(-\chi^2_{\rm scan}/2)$.

The $\chi^2$ of DM relic density is described by the Gaussian distribution
\begin{equation}
    \chi^2_{\Omega h^2} = \left( \frac{\mu_t-\mu_0}{\sqrt{\sigma^2_{\rm theo}+\sigma^2_{\rm exp}}} \right)^2 \, ,
\end{equation}
where $\mu_t$ is the predicted theoretical value we obtained, and $\mu_0$ is the experimental central value. We introduce a 10\% theoretical uncertainty, $\sigma^2_{\rm theo} = 0.1 \mu_t$, to account for  uncertainties from the Boltzmann equation solver and the entropy table in the early Universe. The predicted relic density $\Omega h^2$ is constrained using the PLANCK 2018 data \cite{Planck:2018vyg}, which reports a central value with statistical error as $\Omega h^2 = 0.118 \pm 0.002$.

The estimation of $\chi^2$ for the DM-nucleus spin-independent and spin-dependent cross section is given by:
\begin{equation}
    \chi^2_{\rm DD} = \left( \frac{\sigma_{\rm DD}}{\sigma_{\rm DD}^0/1.64} \right)^2  ,
    \label{eq:DD}
\end{equation}
where $\sigma_{\rm DD}$ represents the predicted theoretical value obtained from \texttt{MicroMEGAS-5.3.41}, and $\sigma_{\rm DD}^0$ denotes the upper limits of the cross sections for a given DM mass at the 90\% confidence level. These limits are provided by experiments such as DarkSide-50 \cite{DarkSide:2018bpj}, CRESST-III \cite{CRESST:2019jnq}, PICO-60 \cite{PICO:2019vsc}, PandaX-4T \cite{PandaX-4T:2021bab}, and Xenon1T \cite{XENON:2018voc}.

\begin{figure}
    \centering
    \includegraphics[width=0.8\textwidth]{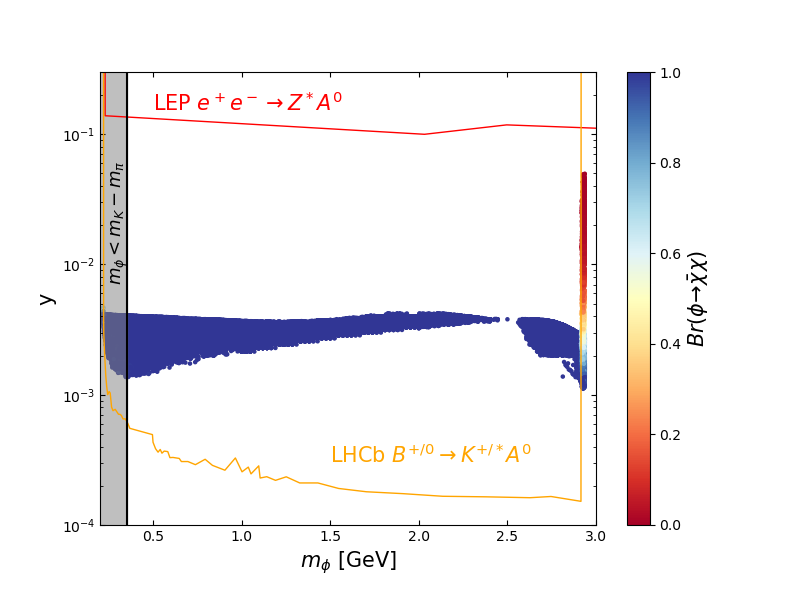}
    \caption{Survived sample points with the constraints $\chi^2_{\rm tot} < 6$ and $Br[B\rightarrow K + \bar \chi \chi] < 2\sigma$. The colorbar represents the branching fraction of $\phi \to \bar \chi \chi$ process. The red and orange solid lines are the upper limits of experimental constraints from LEP \cite{L3:1996ome} and LHCb \cite{1508.04094, 1612.07818}, respectively. The gray shaded region is excluded by $Br[K^+\to \pi^+ \phi]$ data.}
    \label{fig:final}
\end{figure}

Apart form the constraints from relic density and DD implemented during the scanning process, we also considered following phenomenological implications of the model which may challenge the possible surviving phase space:

\begin{itemize}
    \item The model Eq. \eqref{eq:lagrangian} also generates processes like $b\to d \phi$ and $s\to d \phi$, which contribute to $B^0 \to {\rm invisible}$, $B^+\to \pi^+ \cancel{E}$, and $K^+ \to \pi^+ \cancel{E}$, by changing the appropriate quark fields, CKM matrix elements or form factors. 

    Although matrix element $\langle0|\bar bd|B\rangle$ vanishes \cite{1005.1277}, $\langle0|\bar b \gamma_5 d|B\rangle = -i \frac{f_B m_B^2}{m_b+m_d}$ contribute to the decay width of  the process $B^0 \to {\rm invisible}$, which is 
    \begin{equation}
        \Gamma_{B^0 \to \chi \bar \chi} \simeq 
        \left(\frac{y \kappa G_F m_t V_{td}^* V_{tb}}{v}\right)^2 \frac{f_B^2 m_B^5}{16\pi (m_b+m_d)^2} \sqrt{1-\frac{4m_\chi^2}{m_B^2}} \frac{1}{(m_B^2-m_\phi^2)^2} \ .
    \end{equation}
    When $2 m_\chi = m_\phi < 3$  GeV and $\kappa=1$, branching ratio is $Br[B^0 \to {\rm invisible}] \lesssim 3.8 \times 10^{-5} y^2$. Therefore the experimental upper limit $Br[B^0 \to {\rm invisible}] < 7.8 \times 10^{-5}$ \cite{2004.03826} is easily satisfied.

    For the process $B^+\to \pi^+ \phi$, the corresponding decay width can be obtained by substituting the CKM matrix elements $V_{ts}$ with $V_{td}$ and replacing the kaon mass $m_K$ with and pion mass $m_\pi$ in Eq. \eqref{eq:width}. Given that $V_{td} \simeq 0.2 V_{ts}$ \cite{ParticleDataGroup:2022pth}, the decay width $\Gamma_{B\to \pi \phi}$ is approximately $ 0.04 \Gamma_{B\to K \phi}$. Since the experimental upper limit $Br[B^+ \to \pi^+ \cancel{E}] < 6.4\times 10^{-6}$ \cite{2103.12921} is comparable to $Br[B^+\to K^+ \phi]$, the $B^+\to \pi^+ \phi$ process does not impose additional constraints.

    Similarly, the decay width for $K^+ \to \pi^+ \phi$ is determined by replacing $V_{tb}$, bottom quark mass $m_b$, $m_B$, $m_K$ and the form factor $f_0(q^2)$ with $V_{td}$, strange quark mass $m_s$, $m_K$, $m_\pi$ and the $K \to \pi$ form factor $f_0^{K^+}(q^2)$, respectively. The form factor $f_0^{K^+}(q^2) = f_+^{K^+}(0) (1+\lambda_0 q^2/m_\pi^2)$, where $f_+^{K^+}(0) = 09778$ and $\lambda_0 = 0.01338$ \cite{0705.2025,2405.06742}. For a $y \sim 10^{-3}$, the branching ratio $Br[K^+ \to \pi^+ \phi]$ is at the order of $10^{-6}$, which significantly exceeding the experimental upper limit of $Br[K^+ \to \pi^+ \cancel{E}] \lesssim 10^{-11}$ \cite{2103.15389}. Therefore, $m_\phi$ should not be lighter than $m_K-m_\pi$, as illustrated in the gray shaded region of FIG. \ref{fig:final}.

    \item{(ii)} The scalar mediator $\phi$ couple to the all SM fermions, thereby influencing rare leptonic decays such as $b\to (s,d)l^+l^-$ and $s\to dl^+l^-$, $l=e,\mu$. Additionally, hadronic decays like $b\to (s,d)q\bar q$ and $s\to dq\bar q$ can also constrain the parameter space. Several studies have investigated the decay of scalar mediators to SM fermion pairs\cite{1310.6752,1512.04119,1706.01920,1809.01876,2112.11852,2306.02830}, providing constraints on the coupling parameter as a function of  $m_\phi$. These constraints are critical in defining the viable parameter space of our model.

    \textit{A distinct aspect of this paper, compared to previous works, is that the branching fraction of $\phi$ decay to all possible SM particles, denoted as $Br(\phi \rightarrow {\rm SM\ SM})$, is not necessarily unity.} Consequently, we must adjust the limits derived from experiments such as LEP \cite{L3:1996ome}, Belle \cite{Belle:2021rcl}, BESIII \cite{2109.12625}, LHCb \cite{1508.04094, 1612.07818}, NA62 \cite{2011.11329, 2103.15389, 2010.07644}, KTeV \cite{1809.01876, hep-ex/0309072, 0805.0031}, CHARM \cite{1809.01876, 1310.8042, 0912.0390, 1310.6752, 1504.04855} and PS191 \cite{2105.11102}. If we posit $Br(\phi \rightarrow {\rm SM\ SM}) = f_{\rm SM}$ for a specific parameter set where $y=y_0$ and other parameters are fixed, the equivalent $y$ should be recalculated as
    \begin{equation}
        y^\prime=\left(y_0^4/f_{\rm SM}\right)^{1/4} ,
        \label{eq:ratio}
    \end{equation}
    assuming that SM particles are the sole decay products. Therefore, the exclusion limits becomes more robust if the branching fractions to the SM final states are less than one.

    The estimation of $\chi^2$ for the $y$ can be analogized to the expression for $\chi^2_{\rm DD}$ given by Eq. \eqref{eq:DD}. By integrating the relationship defined in Eq. \eqref{eq:ratio}, we can formulate $\chi^2_y$ as follows:

    \begin{equation}
        \chi^2_y = \left( \frac{y}{f_{\rm SM}^{1/4}y^{90\%}/1.64} \right)^2 ,
    \end{equation}

    where $y$ represents the coupling parameter at our sample points, and $y^{90\%}$ denotes the experimental upper limits of $y$ for a given $m_\phi$ at the 90\% confidence level from the referenced experiments. This formulation ensures that $\chi^2_y$ captures the maximum deviation of the observed coupling from its experimental limits, weighted by the branching fraction to SM particles.

    In FIG. \ref{fig:final}, we present two primary constraint lines from LEP \cite{L3:1996ome} and LHCb \cite{1508.04094, 1612.07818}. These constraints rule out ($\chi^2_y > 4$) the sample points with SM final states, thereby only the DM final decay processes are survived in this region. 

    \item{(iii)} The model generates decaying channels of Higgs and weak bosons at the tree level if kinematically allowed, $H/Z \to f\bar f \phi$ and $W\to f \bar f^\prime \phi$, as well as the top-quark decay $t\to bW\phi$. However, these tree-level diagrams are significantly suppressed by a factor of about $(\frac{y m_f}{v})^2$. Given that the largest possible value for $m_f$ in this factor is the mass of bottom quark, and considering $y \simeq 10^{-3}$, the contributions from these processes to the decay widths of $H$, $Z$, $W$, and $t$ are negligible.

    \item{(iv)} Additionally, our model also generates Higgs and Z boson invisible decays at the one loop level, $H\to \phi \phi$ and $Z\to \phi \phi$. The suppression for these decay processes can be quantified as approximately $(\frac{ym_t}{v}\frac{m_t}{m_b})^2\frac{1}{m_H^4}$ for the Higgs and $(\frac{ym_t}{v})^2\frac{1}{m_Z^4}$ for the $Z$ boson. These suppression factors render the contributions to the decay widths significantly smaller than the experimental uncertainties associated with these processes.

\end{itemize}

\section{Results and discussion} \label{sec:results}

In the FIG. \ref{fig:final}, we present sample plots where total $\chi^2$ value, $\chi^2_{\rm tot} = \chi^2_{\rm scan} + \chi^2_y$, is less than 6, with the minimum value of $\chi^2_{\rm tot} \simeq 0$. These plots also ensure that the branching fraction $Br[B\rightarrow K + \bar \chi \chi]$ falls within the $2\sigma$ range. The relevant constraint lines from LEP \cite{L3:1996ome} and LHCb \cite{1508.04094, 1612.07818} are also shown. It is evident from the figure that, in regions above the experimental upper limits, the branching fraction to the DM pair must be very close to one for these sample points to survive. 

SM final state particles can be part of $\phi$ decay channels below the experimental exclusion lines. The closer to the upper limit, the bigger $Br[\phi \rightarrow SM\ SM]$ gets as indicated in Eq. \eqref{eq:ratio}. These sample points with nonzero $Br[\phi \rightarrow SM\ SM]$ are located at the charmonium resonant area. Actually, this narrow resonance area was vetoed by excluding the region to avoid contamination, and no search is performed in the original Ref.  \cite{1508.04094} .

There is no sample point survive when $m_\phi \lesssim 2m_\mu$, since where only $\bar \chi \chi \rightarrow e^+ e^-$ channel allowed and corresponding annihilation cross section is too small even if one introduces the resonant enhancement. However, constraint from $Br[K^+\to \pi^+ \phi]$ data is even stronger, and push the lower limit of $m_\phi$ to the $m_k - m_\pi$. When annihilation process $\bar \chi \chi \rightarrow \bar c c$ is closed, DM relic density is on the high side, and that is the reason why there is a gap near $m_\phi \simeq 2.5$ GeV.

Note that if the lifetime of $\phi$ particle is long enough to escape from the detector before decaying, $\phi$ acts as the missing energy regardless of the decay products. However, to match with DM relic density and $Br[B\rightarrow K+\phi]$, both $\kappa$ and $y$ can not be too small. In our sample points in FIG. \ref{fig:final}, maximum lifetime of the $\phi$ particle is of the order of $\sim 10^{-14}$s, thus it can be seen as prompt decay in the detector. For the long-lived $\phi$ and displaced vertex, one can resort to the Refs. \cite{1809.01876,2306.02830,Filimonova:2019tuy}, where the exclusion capability may differ.

On the other hand, if $m_\chi > m_\phi/2$, although the decay products of $\phi$ are only the SM particles, long-lived $\phi$ relative to the detector size behaves like missing energy. When $y \simeq 3 \times 10^{-3}$ and $m_\phi < 2m_\mu$, the decay width $\Gamma (\phi \rightarrow e^+ e^-)$ can be less than $10^{-15}$ GeV to be invisible. Additionally, $m_\phi$ must be heavier than electron pair mass to ensure decay prior to the big bang nucleosynthesis. In this scenario, heavy DM can easily satisfy the observed relic density \cite{1310.6752,1809.01876}. Such a DM with light scalar mediator that satisfies the $B$-meson anomaly constraints is predicted to have a sizable direct detection cross section~\cite{Elor:2021swj,Bhattiprolu:2022sdd}, and this makes it ready to test in the near future experiments.

\section{Conclusion}\label{sec:conclud}
In this paper, we have explored a simple model involving a light mediator $\phi$ and a light DM particle $\chi$, inspired by the reinterpretation of recent Belle II results from the $B\rightarrow K + \bar \nu \nu$ measurement into the two-body decay $B\rightarrow K + \phi$. Through a comprehensive analysis of the parameter space, supported by experimental results, we have identified that when $2m_\chi < m_\phi$-allowing for the decay $\phi \rightarrow \bar \chi \chi$-the mass of $\phi$ must be heavier than that of the muon pair. Moreover, constraints from the $Br[K^+\to \pi^+ \phi]$ data extend the lower mass limit of $\phi$ to approximately 0.35 GeV, equivalent to $m_K -m_\pi$. In addition, the branching fraction of $\phi$ to the DM pair must be very close to the one to escape the constraints from LHCb, except in a narrow region near $m_\phi \simeq 3$ GeV, where $\phi$ can decay to SM particles partially. When $m_\chi > m_\phi/2$, DM easily have correct relic density, $\phi$ should play the role of missing energy in the detector, and small width of $\phi$ requires $m_\phi < 2 m_\mu$. In other word, $\phi$ can only decay to electron pair. The model in this work is rather simple and includes small number of parameters, and its predictions are within the reach of current or near future experiments, which increases testability of the model.

\section*{Acknowledgment}
M.A. is supported by the National Natural Science Foundation of the People’s Republic of China (No. 12303002) and Tianchi Doctoral Project of Xinjiang Uygur autonomous region of China.
Y.R. is supported by the Natural Science Foundation of Xinjiang Uygur Autonomous Region of China under Grant No.~2022D01C52.

\appendix
\section{$B \to K$ form factors}\label{appendix:FF}

We adopted two different form factor schemes for the stability of our result. The first one is from Ref. \cite{2301.06990}, which is the combination of latest lattice result from Ref. \cite{2207.12468} and the previous lattice one in Ref. \cite{1509.06235}. The second one is outlined in Ref. \cite{2304.12837}, which is the combination of Ref. \cite{2207.12468} with the latest Light-Cone Sum Rule (LCSR) result of Ref. \cite{Gubernari:2018wyi}.

(1) In Ref. \cite{2301.06990}, parametrization for the $B\to K$ form factor is from FLAG \cite{2111.09849}

\begin{equation}
    f_0(q^2) = \frac{1}{1-q^2/M_0^2} [a_0 + a_1 z(q^2)],
\end{equation}
with$M_0 = 5.711$ GeV, and
\begin{equation}
    z(q^2) = \frac{\sqrt{t_+ -q^2}-\sqrt{t_+ -t_0}}{\sqrt{t_+ -q^2}+\sqrt{t_+ -t_0}},
\end{equation}
where $t_+ = (m_B+m_K)^2$ and $t_0 = (m_B+m_K)(\sqrt{m_B}-\sqrt{m_K})^2$. Coefficients $a_0$, $a_1$and their correlation matrix element are 0.2939(36), 0.227(40) and 0.4568, respectively.

(2) In Ref. \cite{2304.12837}, they adopted parametrization \cite{1503.05534}

\begin{equation}
    f_0(q^2) = \frac{1}{1-q^2/M_0^2} [a_0 + a_1 (z(q^2)-z(0)) + a_2 (z(q^2)-z(0))^2],
\end{equation}
with coefficients and correlation matrix are given as in TABLE \ref{tab:coeff}.

\begin{table}[h]
    \centering
    \begin{tabular}{ccc}
    \hline
         $a_0$  &   $a_1$ & $a_2$\\
    \hline
    0.3233(67)  &0.214(57)& -0.12(13) \\
    \hline
    \hline
                & 0.8083  & 0.5481 \\
                &         & 0.9074 \\
    \hline
    
    \end{tabular}
\caption{Fit results of the coefficients with uncertainties and the correlation matrix \cite{2304.12837}.}
\label{tab:coeff}
\end{table}

\begin{figure}[htbp]
    \centering
    \includegraphics[width=0.6\textwidth]{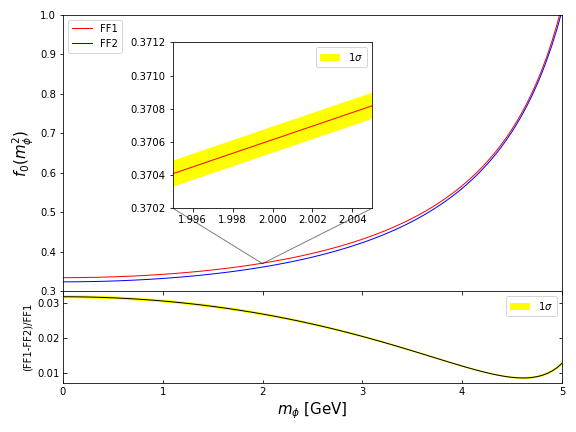}
    \caption{The form factors and relative differences as a function of $m_\phi$. FF1 is for the fitted results from Ref. \cite{2301.06990}, while FF2 is for the Ref. \cite{2304.12837}.}
    \label{fig:ff}
\end{figure}

For a quantitative comparison between two form factor schemes described above, we draw the FIG. \ref{fig:ff}. The gap between two schemes become larger for small $m_\phi$, but relative difference at most about 3\%  as shown in the subfigure below. The 1 $\sigma$ uncertainty range for the form factors are very narrow. For the clear illustration, enlarged a small area near $m_\phi \sim 2$ GeV for the first form factor scheme.

\bibliography{refs}

\begin{thebibliography}{68}%
\makeatletter
\providecommand \@ifxundefined [1]{%
 \@ifx{#1\undefined}
}%
\providecommand \@ifnum [1]{%
 \ifnum #1\expandafter \@firstoftwo
 \else \expandafter \@secondoftwo
 \fi
}%
\providecommand \@ifx [1]{%
 \ifx #1\expandafter \@firstoftwo
 \else \expandafter \@secondoftwo
 \fi
}%
\providecommand \natexlab [1]{#1}%
\providecommand \enquote  [1]{``#1''}%
\providecommand \bibnamefont  [1]{#1}%
\providecommand \bibfnamefont [1]{#1}%
\providecommand \citenamefont [1]{#1}%
\providecommand \href@noop [0]{\@secondoftwo}%
\providecommand \href [0]{\begingroup \@sanitize@url \@href}%
\providecommand \@href[1]{\@@startlink{#1}\@@href}%
\providecommand \@@href[1]{\endgroup#1\@@endlink}%
\providecommand \@sanitize@url [0]{\catcode `\\12\catcode `\$12\catcode
  `\&12\catcode `\#12\catcode `\^12\catcode `\_12\catcode `\%12\relax}%
\providecommand \@@startlink[1]{}%
\providecommand \@@endlink[0]{}%
\providecommand \url  [0]{\begingroup\@sanitize@url \@url }%
\providecommand \@url [1]{\endgroup\@href {#1}{\urlprefix }}%
\providecommand \urlprefix  [0]{URL }%
\providecommand \Eprint [0]{\href }%
\providecommand \doibase [0]{http://dx.doi.org/}%
\providecommand \selectlanguage [0]{\@gobble}%
\providecommand \bibinfo  [0]{\@secondoftwo}%
\providecommand \bibfield  [0]{\@secondoftwo}%
\providecommand \translation [1]{[#1]}%
\providecommand \BibitemOpen [0]{}%
\providecommand \bibitemStop [0]{}%
\providecommand \bibitemNoStop [0]{.\EOS\space}%
\providecommand \EOS [0]{\spacefactor3000\relax}%
\providecommand \BibitemShut  [1]{\csname bibitem#1\endcsname}%
\let\auto@bib@innerbib\@empty
\bibitem [{\citenamefont {Glashow}\ \emph {et~al.}(1970)\citenamefont
  {Glashow}, \citenamefont {Iliopoulos},\ and\ \citenamefont
  {Maiani}}]{Glashow:1970gm}%
  \BibitemOpen
  \bibfield  {author} {\bibinfo {author} {\bibfnamefont {S.~L.}\ \bibnamefont
  {Glashow}}, \bibinfo {author} {\bibfnamefont {J.}~\bibnamefont {Iliopoulos}},
  \ and\ \bibinfo {author} {\bibfnamefont {L.}~\bibnamefont {Maiani}},\ }\href
  {\doibase 10.1103/PhysRevD.2.1285} {\bibfield  {journal} {\bibinfo  {journal}
  {Phys. Rev. D}\ }\textbf {\bibinfo {volume} {2}},\ \bibinfo {pages} {1285}
  (\bibinfo {year} {1970})}\BibitemShut {NoStop}%
\bibitem [{\citenamefont {Parrott}\ \emph
  {et~al.}(2023{\natexlab{a}})\citenamefont {Parrott}, \citenamefont
  {Bouchard},\ and\ \citenamefont {Davies}}]{Parrott:2022zte}%
  \BibitemOpen
  \bibfield  {author} {\bibinfo {author} {\bibfnamefont {W.~G.}\ \bibnamefont
  {Parrott}}, \bibinfo {author} {\bibfnamefont {C.}~\bibnamefont {Bouchard}}, \
  and\ \bibinfo {author} {\bibfnamefont {C.~T.~H.}\ \bibnamefont {Davies}}
  (\bibinfo {collaboration} {HPQCD}),\ }\href {\doibase
  10.1103/PhysRevD.107.014511} {\bibfield  {journal} {\bibinfo  {journal}
  {Phys. Rev. D}\ }\textbf {\bibinfo {volume} {107}},\ \bibinfo {pages}
  {014511} (\bibinfo {year} {2023}{\natexlab{a}})},\ \bibinfo {note} {[Erratum:
  Phys.Rev.D 107, 119903 (2023)]},\ \Eprint {http://arxiv.org/abs/2207.13371}
  {arXiv:2207.13371 [hep-ph]} \BibitemShut {NoStop}%
\bibitem [{\citenamefont {Adachi}\ \emph
  {et~al.}(2023{\natexlab{a}})\citenamefont {Adachi} \emph
  {et~al.}}]{Belle-II:2023esi}%
  \BibitemOpen
  \bibfield  {author} {\bibinfo {author} {\bibfnamefont {I.}~\bibnamefont
  {Adachi}} \emph {et~al.} (\bibinfo {collaboration} {Belle-II}),\ }\href@noop
  {} {\  (\bibinfo {year} {2023}{\natexlab{a}})},\ \Eprint
  {http://arxiv.org/abs/2311.14647} {arXiv:2311.14647 [hep-ex]} \BibitemShut
  {NoStop}%
\bibitem [{\citenamefont {Altmannshofer}\ \emph {et~al.}(2024)\citenamefont
  {Altmannshofer}, \citenamefont {Crivellin}, \citenamefont {Haigh},
  \citenamefont {Inguglia},\ and\ \citenamefont
  {Martin~Camalich}}]{Altmannshofer:2023hkn}%
  \BibitemOpen
  \bibfield  {author} {\bibinfo {author} {\bibfnamefont {W.}~\bibnamefont
  {Altmannshofer}}, \bibinfo {author} {\bibfnamefont {A.}~\bibnamefont
  {Crivellin}}, \bibinfo {author} {\bibfnamefont {H.}~\bibnamefont {Haigh}},
  \bibinfo {author} {\bibfnamefont {G.}~\bibnamefont {Inguglia}}, \ and\
  \bibinfo {author} {\bibfnamefont {J.}~\bibnamefont {Martin~Camalich}},\
  }\href {\doibase 10.1103/PhysRevD.109.075008} {\bibfield  {journal} {\bibinfo
   {journal} {Phys. Rev. D}\ }\textbf {\bibinfo {volume} {109}},\ \bibinfo
  {pages} {075008} (\bibinfo {year} {2024})},\ \Eprint
  {http://arxiv.org/abs/2311.14629} {arXiv:2311.14629 [hep-ph]} \BibitemShut
  {NoStop}%
\bibitem [{\citenamefont {Martin~Camalich}\ \emph {et~al.}(2020)\citenamefont
  {Martin~Camalich}, \citenamefont {Pospelov}, \citenamefont {Vuong},
  \citenamefont {Ziegler},\ and\ \citenamefont
  {Zupan}}]{MartinCamalich:2020dfe}%
  \BibitemOpen
  \bibfield  {author} {\bibinfo {author} {\bibfnamefont {J.}~\bibnamefont
  {Martin~Camalich}}, \bibinfo {author} {\bibfnamefont {M.}~\bibnamefont
  {Pospelov}}, \bibinfo {author} {\bibfnamefont {P.~N.~H.}\ \bibnamefont
  {Vuong}}, \bibinfo {author} {\bibfnamefont {R.}~\bibnamefont {Ziegler}}, \
  and\ \bibinfo {author} {\bibfnamefont {J.}~\bibnamefont {Zupan}},\ }\href
  {\doibase 10.1103/PhysRevD.102.015023} {\bibfield  {journal} {\bibinfo
  {journal} {Phys. Rev. D}\ }\textbf {\bibinfo {volume} {102}},\ \bibinfo
  {pages} {015023} (\bibinfo {year} {2020})},\ \Eprint
  {http://arxiv.org/abs/2002.04623} {arXiv:2002.04623 [hep-ph]} \BibitemShut
  {NoStop}%
\bibitem [{\citenamefont {Ferber}\ \emph {et~al.}(2023)\citenamefont {Ferber},
  \citenamefont {Filimonova}, \citenamefont {Sch\"afer},\ and\ \citenamefont
  {Westhoff}}]{Ferber:2022rsf}%
  \BibitemOpen
  \bibfield  {author} {\bibinfo {author} {\bibfnamefont {T.}~\bibnamefont
  {Ferber}}, \bibinfo {author} {\bibfnamefont {A.}~\bibnamefont {Filimonova}},
  \bibinfo {author} {\bibfnamefont {R.}~\bibnamefont {Sch\"afer}}, \ and\
  \bibinfo {author} {\bibfnamefont {S.}~\bibnamefont {Westhoff}},\ }\href
  {\doibase 10.1007/JHEP04(2023)131} {\bibfield  {journal} {\bibinfo  {journal}
  {JHEP}\ }\textbf {\bibinfo {volume} {04}},\ \bibinfo {pages} {131} (\bibinfo
  {year} {2023})},\ \Eprint {http://arxiv.org/abs/2201.06580} {arXiv:2201.06580
  [hep-ph]} \BibitemShut {NoStop}%
\bibitem [{\citenamefont {Srednicki}\ \emph {et~al.}(1988)\citenamefont
  {Srednicki}, \citenamefont {Watkins},\ and\ \citenamefont
  {Olive}}]{Srednicki:1988ce}%
  \BibitemOpen
  \bibfield  {author} {\bibinfo {author} {\bibfnamefont {M.}~\bibnamefont
  {Srednicki}}, \bibinfo {author} {\bibfnamefont {R.}~\bibnamefont {Watkins}},
  \ and\ \bibinfo {author} {\bibfnamefont {K.~A.}\ \bibnamefont {Olive}},\
  }\href {\doibase 10.1016/0550-3213(88)90099-5} {\bibfield  {journal}
  {\bibinfo  {journal} {Nucl. Phys. B}\ }\textbf {\bibinfo {volume} {310}},\
  \bibinfo {pages} {693} (\bibinfo {year} {1988})}\BibitemShut {NoStop}%
\bibitem [{\citenamefont {Ellis}\ \emph {et~al.}(1984)\citenamefont {Ellis},
  \citenamefont {Hagelin}, \citenamefont {Nanopoulos}, \citenamefont {Olive},\
  and\ \citenamefont {Srednicki}}]{Ellis:1983ew}%
  \BibitemOpen
  \bibfield  {author} {\bibinfo {author} {\bibfnamefont {J.~R.}\ \bibnamefont
  {Ellis}}, \bibinfo {author} {\bibfnamefont {J.~S.}\ \bibnamefont {Hagelin}},
  \bibinfo {author} {\bibfnamefont {D.~V.}\ \bibnamefont {Nanopoulos}},
  \bibinfo {author} {\bibfnamefont {K.~A.}\ \bibnamefont {Olive}}, \ and\
  \bibinfo {author} {\bibfnamefont {M.}~\bibnamefont {Srednicki}},\ }\href
  {\doibase 10.1016/0550-3213(84)90461-9} {\bibfield  {journal} {\bibinfo
  {journal} {Nucl. Phys. B}\ }\textbf {\bibinfo {volume} {238}},\ \bibinfo
  {pages} {453} (\bibinfo {year} {1984})}\BibitemShut {NoStop}%
\bibitem [{\citenamefont {Lee}\ and\ \citenamefont
  {Weinberg}(1977)}]{Lee:1977ua}%
  \BibitemOpen
  \bibfield  {author} {\bibinfo {author} {\bibfnamefont {B.~W.}\ \bibnamefont
  {Lee}}\ and\ \bibinfo {author} {\bibfnamefont {S.}~\bibnamefont {Weinberg}},\
  }\href {\doibase 10.1103/PhysRevLett.39.165} {\bibfield  {journal} {\bibinfo
  {journal} {Phys. Rev. Lett.}\ }\textbf {\bibinfo {volume} {39}},\ \bibinfo
  {pages} {165} (\bibinfo {year} {1977})}\BibitemShut {NoStop}%
\bibitem [{\citenamefont {Pospelov}\ \emph {et~al.}(2008)\citenamefont
  {Pospelov}, \citenamefont {Ritz},\ and\ \citenamefont
  {Voloshin}}]{Pospelov:2007mp}%
  \BibitemOpen
  \bibfield  {author} {\bibinfo {author} {\bibfnamefont {M.}~\bibnamefont
  {Pospelov}}, \bibinfo {author} {\bibfnamefont {A.}~\bibnamefont {Ritz}}, \
  and\ \bibinfo {author} {\bibfnamefont {M.~B.}\ \bibnamefont {Voloshin}},\
  }\href {\doibase 10.1016/j.physletb.2008.02.052} {\bibfield  {journal}
  {\bibinfo  {journal} {Phys. Lett. B}\ }\textbf {\bibinfo {volume} {662}},\
  \bibinfo {pages} {53} (\bibinfo {year} {2008})},\ \Eprint
  {http://arxiv.org/abs/0711.4866} {arXiv:0711.4866 [hep-ph]} \BibitemShut
  {NoStop}%
\bibitem [{\citenamefont {He}\ \emph {et~al.}(2023)\citenamefont {He},
  \citenamefont {Ma},\ and\ \citenamefont {Valencia}}]{He:2022ljo}%
  \BibitemOpen
  \bibfield  {author} {\bibinfo {author} {\bibfnamefont {X.-G.}\ \bibnamefont
  {He}}, \bibinfo {author} {\bibfnamefont {X.-D.}\ \bibnamefont {Ma}}, \ and\
  \bibinfo {author} {\bibfnamefont {G.}~\bibnamefont {Valencia}},\ }\href
  {\doibase 10.1007/JHEP03(2023)037} {\bibfield  {journal} {\bibinfo  {journal}
  {JHEP}\ }\textbf {\bibinfo {volume} {03}},\ \bibinfo {pages} {037} (\bibinfo
  {year} {2023})},\ \Eprint {http://arxiv.org/abs/2209.05223} {arXiv:2209.05223
  [hep-ph]} \BibitemShut {NoStop}%
\bibitem [{\citenamefont {Li}\ \emph {et~al.}(2021{\natexlab{a}})\citenamefont
  {Li}, \citenamefont {Wang}, \citenamefont {Zhang},\ and\ \citenamefont
  {Wang}}]{Li:2021sqe}%
  \BibitemOpen
  \bibfield  {author} {\bibinfo {author} {\bibfnamefont {G.}~\bibnamefont
  {Li}}, \bibinfo {author} {\bibfnamefont {T.}~\bibnamefont {Wang}}, \bibinfo
  {author} {\bibfnamefont {J.-B.}\ \bibnamefont {Zhang}}, \ and\ \bibinfo
  {author} {\bibfnamefont {G.-L.}\ \bibnamefont {Wang}},\ }\href {\doibase
  10.1140/epjc/s10052-021-09333-z} {\bibfield  {journal} {\bibinfo  {journal}
  {Eur. Phys. J. C}\ }\textbf {\bibinfo {volume} {81}},\ \bibinfo {pages} {564}
  (\bibinfo {year} {2021}{\natexlab{a}})},\ \Eprint
  {http://arxiv.org/abs/2103.12921} {arXiv:2103.12921 [hep-ph]} \BibitemShut
  {NoStop}%
\bibitem [{\citenamefont {Kamenik}\ and\ \citenamefont
  {Smith}(2012)}]{Kamenik:2011vy}%
  \BibitemOpen
  \bibfield  {author} {\bibinfo {author} {\bibfnamefont {J.~F.}\ \bibnamefont
  {Kamenik}}\ and\ \bibinfo {author} {\bibfnamefont {C.}~\bibnamefont
  {Smith}},\ }\href {\doibase 10.1007/JHEP03(2012)090} {\bibfield  {journal}
  {\bibinfo  {journal} {JHEP}\ }\textbf {\bibinfo {volume} {03}},\ \bibinfo
  {pages} {090} (\bibinfo {year} {2012})},\ \Eprint
  {http://arxiv.org/abs/1111.6402} {arXiv:1111.6402 [hep-ph]} \BibitemShut
  {NoStop}%
\bibitem [{\citenamefont {Patt}\ and\ \citenamefont
  {Wilczek}(2006)}]{Patt:2006fw}%
  \BibitemOpen
  \bibfield  {author} {\bibinfo {author} {\bibfnamefont {B.}~\bibnamefont
  {Patt}}\ and\ \bibinfo {author} {\bibfnamefont {F.}~\bibnamefont {Wilczek}},\
  }\href@noop {} {\  (\bibinfo {year} {2006})},\ \Eprint
  {http://arxiv.org/abs/hep-ph/0605188} {arXiv:hep-ph/0605188} \BibitemShut
  {NoStop}%
\bibitem [{\citenamefont {Winkler}(2019)}]{1809.01876}%
  \BibitemOpen
  \bibfield  {author} {\bibinfo {author} {\bibfnamefont {M.~W.}\ \bibnamefont
  {Winkler}},\ }\href {\doibase 10.1103/PhysRevD.99.015018} {\bibfield
  {journal} {\bibinfo  {journal} {Phys. Rev. D}\ }\textbf {\bibinfo {volume}
  {99}},\ \bibinfo {pages} {015018} (\bibinfo {year} {2019})},\ \Eprint
  {http://arxiv.org/abs/1809.01876} {arXiv:1809.01876 [hep-ph]} \BibitemShut
  {NoStop}%
\bibitem [{\citenamefont {Krnjaic}(2016{\natexlab{a}})}]{Krnjaic:2015mbs}%
  \BibitemOpen
  \bibfield  {author} {\bibinfo {author} {\bibfnamefont {G.}~\bibnamefont
  {Krnjaic}},\ }\href {\doibase 10.1103/PhysRevD.94.073009} {\bibfield
  {journal} {\bibinfo  {journal} {Phys. Rev. D}\ }\textbf {\bibinfo {volume}
  {94}},\ \bibinfo {pages} {073009} (\bibinfo {year} {2016}{\natexlab{a}})},\
  \Eprint {http://arxiv.org/abs/1512.04119} {arXiv:1512.04119 [hep-ph]}
  \BibitemShut {NoStop}%
\bibitem [{\citenamefont {Filimonova}\ \emph {et~al.}(2020)\citenamefont
  {Filimonova}, \citenamefont {Sch\"afer},\ and\ \citenamefont
  {Westhoff}}]{Filimonova:2019tuy}%
  \BibitemOpen
  \bibfield  {author} {\bibinfo {author} {\bibfnamefont {A.}~\bibnamefont
  {Filimonova}}, \bibinfo {author} {\bibfnamefont {R.}~\bibnamefont
  {Sch\"afer}}, \ and\ \bibinfo {author} {\bibfnamefont {S.}~\bibnamefont
  {Westhoff}},\ }\href {\doibase 10.1103/PhysRevD.101.095006} {\bibfield
  {journal} {\bibinfo  {journal} {Phys. Rev. D}\ }\textbf {\bibinfo {volume}
  {101}},\ \bibinfo {pages} {095006} (\bibinfo {year} {2020})},\ \Eprint
  {http://arxiv.org/abs/1911.03490} {arXiv:1911.03490 [hep-ph]} \BibitemShut
  {NoStop}%
\bibitem [{\citenamefont {Kachanovich}\ \emph {et~al.}(2020)\citenamefont
  {Kachanovich}, \citenamefont {Nierste},\ and\ \citenamefont
  {Ni\v{s}and\v{z}i\'c}}]{Kachanovich:2020yhi}%
  \BibitemOpen
  \bibfield  {author} {\bibinfo {author} {\bibfnamefont {A.}~\bibnamefont
  {Kachanovich}}, \bibinfo {author} {\bibfnamefont {U.}~\bibnamefont
  {Nierste}}, \ and\ \bibinfo {author} {\bibfnamefont {I.}~\bibnamefont
  {Ni\v{s}and\v{z}i\'c}},\ }\href {\doibase 10.1140/epjc/s10052-020-8240-z}
  {\bibfield  {journal} {\bibinfo  {journal} {Eur. Phys. J. C}\ }\textbf
  {\bibinfo {volume} {80}},\ \bibinfo {pages} {669} (\bibinfo {year} {2020})},\
  \Eprint {http://arxiv.org/abs/2003.01788} {arXiv:2003.01788 [hep-ph]}
  \BibitemShut {NoStop}%
\bibitem [{\citenamefont {Ovchynnikov}\ \emph {et~al.}(2023)\citenamefont
  {Ovchynnikov}, \citenamefont {Schmidt},\ and\ \citenamefont
  {Schwetz}}]{Ovchynnikov:2023von}%
  \BibitemOpen
  \bibfield  {author} {\bibinfo {author} {\bibfnamefont {M.}~\bibnamefont
  {Ovchynnikov}}, \bibinfo {author} {\bibfnamefont {M.~A.}\ \bibnamefont
  {Schmidt}}, \ and\ \bibinfo {author} {\bibfnamefont {T.}~\bibnamefont
  {Schwetz}},\ }\href {\doibase 10.1140/epjc/s10052-023-11975-0} {\bibfield
  {journal} {\bibinfo  {journal} {Eur. Phys. J. C}\ }\textbf {\bibinfo {volume}
  {83}},\ \bibinfo {pages} {791} (\bibinfo {year} {2023})},\ \Eprint
  {http://arxiv.org/abs/2306.09508} {arXiv:2306.09508 [hep-ph]} \BibitemShut
  {NoStop}%
\bibitem [{\citenamefont {Asadi}\ \emph {et~al.}(2023)\citenamefont {Asadi},
  \citenamefont {Bhattacharya}, \citenamefont {Fraser}, \citenamefont
  {Homiller},\ and\ \citenamefont {Parikh}}]{Asadi:2023ucx}%
  \BibitemOpen
  \bibfield  {author} {\bibinfo {author} {\bibfnamefont {P.}~\bibnamefont
  {Asadi}}, \bibinfo {author} {\bibfnamefont {A.}~\bibnamefont {Bhattacharya}},
  \bibinfo {author} {\bibfnamefont {K.}~\bibnamefont {Fraser}}, \bibinfo
  {author} {\bibfnamefont {S.}~\bibnamefont {Homiller}}, \ and\ \bibinfo
  {author} {\bibfnamefont {A.}~\bibnamefont {Parikh}},\ }\href {\doibase
  10.1007/JHEP10(2023)069} {\bibfield  {journal} {\bibinfo  {journal} {JHEP}\
  }\textbf {\bibinfo {volume} {10}},\ \bibinfo {pages} {069} (\bibinfo {year}
  {2023})},\ \Eprint {http://arxiv.org/abs/2308.01340} {arXiv:2308.01340
  [hep-ph]} \BibitemShut {NoStop}%
\bibitem [{\citenamefont {Athron}\ \emph {et~al.}(2024)\citenamefont {Athron},
  \citenamefont {Martinez},\ and\ \citenamefont {Sierra}}]{Athron:2023hmz}%
  \BibitemOpen
  \bibfield  {author} {\bibinfo {author} {\bibfnamefont {P.}~\bibnamefont
  {Athron}}, \bibinfo {author} {\bibfnamefont {R.}~\bibnamefont {Martinez}}, \
  and\ \bibinfo {author} {\bibfnamefont {C.}~\bibnamefont {Sierra}},\ }\href
  {\doibase 10.1007/JHEP02(2024)121} {\bibfield  {journal} {\bibinfo  {journal}
  {JHEP}\ }\textbf {\bibinfo {volume} {02}},\ \bibinfo {pages} {121} (\bibinfo
  {year} {2024})},\ \Eprint {http://arxiv.org/abs/2308.13426} {arXiv:2308.13426
  [hep-ph]} \BibitemShut {NoStop}%
\bibitem [{\citenamefont {Schmidt-Hoberg}\ \emph
  {et~al.}(2013{\natexlab{a}})\citenamefont {Schmidt-Hoberg}, \citenamefont
  {Staub},\ and\ \citenamefont {Winkler}}]{Schmidt-Hoberg:2013hba}%
  \BibitemOpen
  \bibfield  {author} {\bibinfo {author} {\bibfnamefont {K.}~\bibnamefont
  {Schmidt-Hoberg}}, \bibinfo {author} {\bibfnamefont {F.}~\bibnamefont
  {Staub}}, \ and\ \bibinfo {author} {\bibfnamefont {M.~W.}\ \bibnamefont
  {Winkler}},\ }\href {\doibase 10.1016/j.physletb.2013.11.015} {\bibfield
  {journal} {\bibinfo  {journal} {Phys. Lett. B}\ }\textbf {\bibinfo {volume}
  {727}},\ \bibinfo {pages} {506} (\bibinfo {year} {2013}{\natexlab{a}})},\
  \Eprint {http://arxiv.org/abs/1310.6752} {arXiv:1310.6752 [hep-ph]}
  \BibitemShut {NoStop}%
\bibitem [{\citenamefont {Batell}\ \emph {et~al.}(2011)\citenamefont {Batell},
  \citenamefont {Pospelov},\ and\ \citenamefont {Ritz}}]{Batell:2009jf}%
  \BibitemOpen
  \bibfield  {author} {\bibinfo {author} {\bibfnamefont {B.}~\bibnamefont
  {Batell}}, \bibinfo {author} {\bibfnamefont {M.}~\bibnamefont {Pospelov}}, \
  and\ \bibinfo {author} {\bibfnamefont {A.}~\bibnamefont {Ritz}},\ }\href
  {\doibase 10.1103/PhysRevD.83.054005} {\bibfield  {journal} {\bibinfo
  {journal} {Phys. Rev. D}\ }\textbf {\bibinfo {volume} {83}},\ \bibinfo
  {pages} {054005} (\bibinfo {year} {2011})},\ \Eprint
  {http://arxiv.org/abs/0911.4938} {arXiv:0911.4938 [hep-ph]} \BibitemShut
  {NoStop}%
\bibitem [{\citenamefont {Be\v{c}irevi\'c}\ \emph {et~al.}(2023)\citenamefont
  {Be\v{c}irevi\'c}, \citenamefont {Piazza},\ and\ \citenamefont
  {Sumensari}}]{2301.06990}%
  \BibitemOpen
  \bibfield  {author} {\bibinfo {author} {\bibfnamefont {D.}~\bibnamefont
  {Be\v{c}irevi\'c}}, \bibinfo {author} {\bibfnamefont {G.}~\bibnamefont
  {Piazza}}, \ and\ \bibinfo {author} {\bibfnamefont {O.}~\bibnamefont
  {Sumensari}},\ }\href {\doibase 10.1140/epjc/s10052-023-11388-z} {\bibfield
  {journal} {\bibinfo  {journal} {Eur. Phys. J. C}\ }\textbf {\bibinfo {volume}
  {83}},\ \bibinfo {pages} {252} (\bibinfo {year} {2023})},\ \Eprint
  {http://arxiv.org/abs/2301.06990} {arXiv:2301.06990 [hep-ph]} \BibitemShut
  {NoStop}%
\bibitem [{\citenamefont {Grunwald}\ \emph {et~al.}(2023)\citenamefont
  {Grunwald}, \citenamefont {Hiller}, \citenamefont {Kr\"oninger},\ and\
  \citenamefont {Nollen}}]{2304.12837}%
  \BibitemOpen
  \bibfield  {author} {\bibinfo {author} {\bibfnamefont {C.}~\bibnamefont
  {Grunwald}}, \bibinfo {author} {\bibfnamefont {G.}~\bibnamefont {Hiller}},
  \bibinfo {author} {\bibfnamefont {K.}~\bibnamefont {Kr\"oninger}}, \ and\
  \bibinfo {author} {\bibfnamefont {L.}~\bibnamefont {Nollen}},\ }\href
  {\doibase 10.1007/JHEP11(2023)110} {\bibfield  {journal} {\bibinfo  {journal}
  {JHEP}\ }\textbf {\bibinfo {volume} {11}},\ \bibinfo {pages} {110} (\bibinfo
  {year} {2023})},\ \Eprint {http://arxiv.org/abs/2304.12837} {arXiv:2304.12837
  [hep-ph]} \BibitemShut {NoStop}%
\bibitem [{\citenamefont {McKeen}(2009)}]{McKeen:2008gd}%
  \BibitemOpen
  \bibfield  {author} {\bibinfo {author} {\bibfnamefont {D.}~\bibnamefont
  {McKeen}},\ }\href {\doibase 10.1103/PhysRevD.79.015007} {\bibfield
  {journal} {\bibinfo  {journal} {Phys. Rev. D}\ }\textbf {\bibinfo {volume}
  {79}},\ \bibinfo {pages} {015007} (\bibinfo {year} {2009})},\ \Eprint
  {http://arxiv.org/abs/0809.4787} {arXiv:0809.4787 [hep-ph]} \BibitemShut
  {NoStop}%
\bibitem [{\citenamefont {Gondolo}\ and\ \citenamefont
  {Gelmini}(1991)}]{Gondolo:1990dk}%
  \BibitemOpen
  \bibfield  {author} {\bibinfo {author} {\bibfnamefont {P.}~\bibnamefont
  {Gondolo}}\ and\ \bibinfo {author} {\bibfnamefont {G.}~\bibnamefont
  {Gelmini}},\ }\href {\doibase 10.1016/0550-3213(91)90438-4} {\bibfield
  {journal} {\bibinfo  {journal} {Nucl. Phys. B}\ }\textbf {\bibinfo {volume}
  {360}},\ \bibinfo {pages} {145} (\bibinfo {year} {1991})}\BibitemShut
  {NoStop}%
\bibitem [{\citenamefont {Alloul}\ \emph {et~al.}(2014)\citenamefont {Alloul},
  \citenamefont {Christensen}, \citenamefont {Degrande}, \citenamefont {Duhr},\
  and\ \citenamefont {Fuks}}]{Alloul:2013bka}%
  \BibitemOpen
  \bibfield  {author} {\bibinfo {author} {\bibfnamefont {A.}~\bibnamefont
  {Alloul}}, \bibinfo {author} {\bibfnamefont {N.~D.}\ \bibnamefont
  {Christensen}}, \bibinfo {author} {\bibfnamefont {C.}~\bibnamefont
  {Degrande}}, \bibinfo {author} {\bibfnamefont {C.}~\bibnamefont {Duhr}}, \
  and\ \bibinfo {author} {\bibfnamefont {B.}~\bibnamefont {Fuks}},\ }\href
  {\doibase 10.1016/j.cpc.2014.04.012} {\bibfield  {journal} {\bibinfo
  {journal} {Comput. Phys. Commun.}\ }\textbf {\bibinfo {volume} {185}},\
  \bibinfo {pages} {2250} (\bibinfo {year} {2014})},\ \Eprint
  {http://arxiv.org/abs/1310.1921} {arXiv:1310.1921 [hep-ph]} \BibitemShut
  {NoStop}%
\bibitem [{\citenamefont {Belanger}\ \emph {et~al.}(2011)\citenamefont
  {Belanger}, \citenamefont {Boudjema}, \citenamefont {Brun}, \citenamefont
  {Pukhov}, \citenamefont {Rosier-Lees}, \citenamefont {Salati},\ and\
  \citenamefont {Semenov}}]{Belanger:2010gh}%
  \BibitemOpen
  \bibfield  {author} {\bibinfo {author} {\bibfnamefont {G.}~\bibnamefont
  {Belanger}}, \bibinfo {author} {\bibfnamefont {F.}~\bibnamefont {Boudjema}},
  \bibinfo {author} {\bibfnamefont {P.}~\bibnamefont {Brun}}, \bibinfo {author}
  {\bibfnamefont {A.}~\bibnamefont {Pukhov}}, \bibinfo {author} {\bibfnamefont
  {S.}~\bibnamefont {Rosier-Lees}}, \bibinfo {author} {\bibfnamefont
  {P.}~\bibnamefont {Salati}}, \ and\ \bibinfo {author} {\bibfnamefont
  {A.}~\bibnamefont {Semenov}},\ }\href {\doibase 10.1016/j.cpc.2010.11.033}
  {\bibfield  {journal} {\bibinfo  {journal} {Comput. Phys. Commun.}\ }\textbf
  {\bibinfo {volume} {182}},\ \bibinfo {pages} {842} (\bibinfo {year}
  {2011})},\ \Eprint {http://arxiv.org/abs/1004.1092} {arXiv:1004.1092
  [hep-ph]} \BibitemShut {NoStop}%
\bibitem [{\citenamefont {Acciarri}\ \emph {et~al.}(1996)\citenamefont
  {Acciarri} \emph {et~al.}}]{L3:1996ome}%
  \BibitemOpen
  \bibfield  {author} {\bibinfo {author} {\bibfnamefont {M.}~\bibnamefont
  {Acciarri}} \emph {et~al.} (\bibinfo {collaboration} {L3}),\ }\href {\doibase
  10.1016/0370-2693(96)00987-2} {\bibfield  {journal} {\bibinfo  {journal}
  {Phys. Lett. B}\ }\textbf {\bibinfo {volume} {385}},\ \bibinfo {pages} {454}
  (\bibinfo {year} {1996})}\BibitemShut {NoStop}%
\bibitem [{\citenamefont {Foreman-Mackey}\ \emph {et~al.}(2013)\citenamefont
  {Foreman-Mackey}, \citenamefont {Hogg}, \citenamefont {Lang},\ and\
  \citenamefont {Goodman}}]{Foreman-Mackey:2012any}%
  \BibitemOpen
  \bibfield  {author} {\bibinfo {author} {\bibfnamefont {D.}~\bibnamefont
  {Foreman-Mackey}}, \bibinfo {author} {\bibfnamefont {D.~W.}\ \bibnamefont
  {Hogg}}, \bibinfo {author} {\bibfnamefont {D.}~\bibnamefont {Lang}}, \ and\
  \bibinfo {author} {\bibfnamefont {J.}~\bibnamefont {Goodman}},\ }\href
  {\doibase 10.1086/670067} {\bibfield  {journal} {\bibinfo  {journal} {Publ.
  Astron. Soc. Pac.}\ }\textbf {\bibinfo {volume} {125}},\ \bibinfo {pages}
  {306} (\bibinfo {year} {2013})},\ \Eprint {http://arxiv.org/abs/1202.3665}
  {arXiv:1202.3665 [astro-ph.IM]} \BibitemShut {NoStop}%
\bibitem [{\citenamefont {Aghanim}\ \emph {et~al.}(2020)\citenamefont {Aghanim}
  \emph {et~al.}}]{Planck:2018vyg}%
  \BibitemOpen
  \bibfield  {author} {\bibinfo {author} {\bibfnamefont {N.}~\bibnamefont
  {Aghanim}} \emph {et~al.} (\bibinfo {collaboration} {Planck}),\ }\href
  {\doibase 10.1051/0004-6361/201833910} {\bibfield  {journal} {\bibinfo
  {journal} {Astron. Astrophys.}\ }\textbf {\bibinfo {volume} {641}},\ \bibinfo
  {pages} {A6} (\bibinfo {year} {2020})},\ \bibinfo {note} {[Erratum:
  Astron.Astrophys. 652, C4 (2021)]},\ \Eprint
  {http://arxiv.org/abs/1807.06209} {arXiv:1807.06209 [astro-ph.CO]}
  \BibitemShut {NoStop}%
\bibitem [{\citenamefont {Agnes}\ \emph {et~al.}(2018)\citenamefont {Agnes}
  \emph {et~al.}}]{DarkSide:2018bpj}%
  \BibitemOpen
  \bibfield  {author} {\bibinfo {author} {\bibfnamefont {P.}~\bibnamefont
  {Agnes}} \emph {et~al.} (\bibinfo {collaboration} {DarkSide}),\ }\href
  {\doibase 10.1103/PhysRevLett.121.081307} {\bibfield  {journal} {\bibinfo
  {journal} {Phys. Rev. Lett.}\ }\textbf {\bibinfo {volume} {121}},\ \bibinfo
  {pages} {081307} (\bibinfo {year} {2018})},\ \Eprint
  {http://arxiv.org/abs/1802.06994} {arXiv:1802.06994 [astro-ph.HE]}
  \BibitemShut {NoStop}%
\bibitem [{\citenamefont {Abdelhameed}\ \emph {et~al.}(2019)\citenamefont
  {Abdelhameed} \emph {et~al.}}]{CRESST:2019jnq}%
  \BibitemOpen
  \bibfield  {author} {\bibinfo {author} {\bibfnamefont {A.~H.}\ \bibnamefont
  {Abdelhameed}} \emph {et~al.} (\bibinfo {collaboration} {CRESST}),\ }\href
  {\doibase 10.1103/PhysRevD.100.102002} {\bibfield  {journal} {\bibinfo
  {journal} {Phys. Rev. D}\ }\textbf {\bibinfo {volume} {100}},\ \bibinfo
  {pages} {102002} (\bibinfo {year} {2019})},\ \Eprint
  {http://arxiv.org/abs/1904.00498} {arXiv:1904.00498 [astro-ph.CO]}
  \BibitemShut {NoStop}%
\bibitem [{\citenamefont {Amole}\ \emph {et~al.}(2019)\citenamefont {Amole}
  \emph {et~al.}}]{PICO:2019vsc}%
  \BibitemOpen
  \bibfield  {author} {\bibinfo {author} {\bibfnamefont {C.}~\bibnamefont
  {Amole}} \emph {et~al.} (\bibinfo {collaboration} {PICO}),\ }\href {\doibase
  10.1103/PhysRevD.100.022001} {\bibfield  {journal} {\bibinfo  {journal}
  {Phys. Rev. D}\ }\textbf {\bibinfo {volume} {100}},\ \bibinfo {pages}
  {022001} (\bibinfo {year} {2019})},\ \Eprint
  {http://arxiv.org/abs/1902.04031} {arXiv:1902.04031 [astro-ph.CO]}
  \BibitemShut {NoStop}%
\bibitem [{\citenamefont {Meng}\ \emph {et~al.}(2021)\citenamefont {Meng} \emph
  {et~al.}}]{PandaX-4T:2021bab}%
  \BibitemOpen
  \bibfield  {author} {\bibinfo {author} {\bibfnamefont {Y.}~\bibnamefont
  {Meng}} \emph {et~al.} (\bibinfo {collaboration} {PandaX-4T}),\ }\href
  {\doibase 10.1103/PhysRevLett.127.261802} {\bibfield  {journal} {\bibinfo
  {journal} {Phys. Rev. Lett.}\ }\textbf {\bibinfo {volume} {127}},\ \bibinfo
  {pages} {261802} (\bibinfo {year} {2021})},\ \Eprint
  {http://arxiv.org/abs/2107.13438} {arXiv:2107.13438 [hep-ex]} \BibitemShut
  {NoStop}%
\bibitem [{\citenamefont {Aprile}\ \emph {et~al.}(2018)\citenamefont {Aprile}
  \emph {et~al.}}]{XENON:2018voc}%
  \BibitemOpen
  \bibfield  {author} {\bibinfo {author} {\bibfnamefont {E.}~\bibnamefont
  {Aprile}} \emph {et~al.} (\bibinfo {collaboration} {XENON}),\ }\href
  {\doibase 10.1103/PhysRevLett.121.111302} {\bibfield  {journal} {\bibinfo
  {journal} {Phys. Rev. Lett.}\ }\textbf {\bibinfo {volume} {121}},\ \bibinfo
  {pages} {111302} (\bibinfo {year} {2018})},\ \Eprint
  {http://arxiv.org/abs/1805.12562} {arXiv:1805.12562 [astro-ph.CO]}
  \BibitemShut {NoStop}%
\bibitem [{\citenamefont {Badin}\ and\ \citenamefont
  {Petrov}(2010)}]{1005.1277}%
  \BibitemOpen
  \bibfield  {author} {\bibinfo {author} {\bibfnamefont {A.}~\bibnamefont
  {Badin}}\ and\ \bibinfo {author} {\bibfnamefont {A.~A.}\ \bibnamefont
  {Petrov}},\ }\href {\doibase 10.1103/PhysRevD.82.034005} {\bibfield
  {journal} {\bibinfo  {journal} {Phys. Rev. D}\ }\textbf {\bibinfo {volume}
  {82}},\ \bibinfo {pages} {034005} (\bibinfo {year} {2010})},\ \Eprint
  {http://arxiv.org/abs/1005.1277} {arXiv:1005.1277 [hep-ph]} \BibitemShut
  {NoStop}%
\bibitem [{\citenamefont {Ku}\ \emph {et~al.}(2020)\citenamefont {Ku} \emph
  {et~al.}}]{2004.03826}%
  \BibitemOpen
  \bibfield  {author} {\bibinfo {author} {\bibfnamefont {Y.}~\bibnamefont {Ku}}
  \emph {et~al.} (\bibinfo {collaboration} {Belle}),\ }\href {\doibase
  10.1103/PhysRevD.102.012003} {\bibfield  {journal} {\bibinfo  {journal}
  {Phys. Rev. D}\ }\textbf {\bibinfo {volume} {102}},\ \bibinfo {pages}
  {012003} (\bibinfo {year} {2020})},\ \Eprint
  {http://arxiv.org/abs/2004.03826} {arXiv:2004.03826 [hep-ex]} \BibitemShut
  {NoStop}%
\bibitem [{\citenamefont {Workman}\ \emph {et~al.}(2022)\citenamefont {Workman}
  \emph {et~al.}}]{ParticleDataGroup:2022pth}%
  \BibitemOpen
  \bibfield  {author} {\bibinfo {author} {\bibfnamefont {R.~L.}\ \bibnamefont
  {Workman}} \emph {et~al.} (\bibinfo {collaboration} {Particle Data Group}),\
  }\href {\doibase 10.1093/ptep/ptac097} {\bibfield  {journal} {\bibinfo
  {journal} {PTEP}\ }\textbf {\bibinfo {volume} {2022}},\ \bibinfo {pages}
  {083C01} (\bibinfo {year} {2022})}\BibitemShut {NoStop}%
\bibitem [{\citenamefont {Li}\ \emph {et~al.}(2021{\natexlab{b}})\citenamefont
  {Li}, \citenamefont {Wang}, \citenamefont {Zhang},\ and\ \citenamefont
  {Wang}}]{2103.12921}%
  \BibitemOpen
  \bibfield  {author} {\bibinfo {author} {\bibfnamefont {G.}~\bibnamefont
  {Li}}, \bibinfo {author} {\bibfnamefont {T.}~\bibnamefont {Wang}}, \bibinfo
  {author} {\bibfnamefont {J.-B.}\ \bibnamefont {Zhang}}, \ and\ \bibinfo
  {author} {\bibfnamefont {G.-L.}\ \bibnamefont {Wang}},\ }\href {\doibase
  10.1140/epjc/s10052-021-09333-z} {\bibfield  {journal} {\bibinfo  {journal}
  {Eur. Phys. J. C}\ }\textbf {\bibinfo {volume} {81}},\ \bibinfo {pages} {564}
  (\bibinfo {year} {2021}{\natexlab{b}})},\ \Eprint
  {http://arxiv.org/abs/2103.12921} {arXiv:2103.12921 [hep-ph]} \BibitemShut
  {NoStop}%
\bibitem [{\citenamefont {Mescia}\ and\ \citenamefont
  {Smith}(2007)}]{0705.2025}%
  \BibitemOpen
  \bibfield  {author} {\bibinfo {author} {\bibfnamefont {F.}~\bibnamefont
  {Mescia}}\ and\ \bibinfo {author} {\bibfnamefont {C.}~\bibnamefont {Smith}},\
  }\href {\doibase 10.1103/PhysRevD.76.034017} {\bibfield  {journal} {\bibinfo
  {journal} {Phys. Rev. D}\ }\textbf {\bibinfo {volume} {76}},\ \bibinfo
  {pages} {034017} (\bibinfo {year} {2007})},\ \Eprint
  {http://arxiv.org/abs/0705.2025} {arXiv:0705.2025 [hep-ph]} \BibitemShut
  {NoStop}%
\bibitem [{\citenamefont {Buras}\ \emph {et~al.}(2024)\citenamefont {Buras},
  \citenamefont {Harz},\ and\ \citenamefont {Mojahed}}]{2405.06742}%
  \BibitemOpen
  \bibfield  {author} {\bibinfo {author} {\bibfnamefont {A.~J.}\ \bibnamefont
  {Buras}}, \bibinfo {author} {\bibfnamefont {J.}~\bibnamefont {Harz}}, \ and\
  \bibinfo {author} {\bibfnamefont {M.~A.}\ \bibnamefont {Mojahed}},\
  }\href@noop {} {\  (\bibinfo {year} {2024})},\ \Eprint
  {http://arxiv.org/abs/2405.06742} {arXiv:2405.06742 [hep-ph]} \BibitemShut
  {NoStop}%
\bibitem [{\citenamefont {Cortina~Gil}\ \emph
  {et~al.}(2021{\natexlab{a}})\citenamefont {Cortina~Gil} \emph
  {et~al.}}]{2103.15389}%
  \BibitemOpen
  \bibfield  {author} {\bibinfo {author} {\bibfnamefont {E.}~\bibnamefont
  {Cortina~Gil}} \emph {et~al.} (\bibinfo {collaboration} {NA62}),\ }\href
  {\doibase 10.1007/JHEP06(2021)093} {\bibfield  {journal} {\bibinfo  {journal}
  {JHEP}\ }\textbf {\bibinfo {volume} {06}},\ \bibinfo {pages} {093} (\bibinfo
  {year} {2021}{\natexlab{a}})},\ \Eprint {http://arxiv.org/abs/2103.15389}
  {arXiv:2103.15389 [hep-ex]} \BibitemShut {NoStop}%
\bibitem [{\citenamefont {Schmidt-Hoberg}\ \emph
  {et~al.}(2013{\natexlab{b}})\citenamefont {Schmidt-Hoberg}, \citenamefont
  {Staub},\ and\ \citenamefont {Winkler}}]{1310.6752}%
  \BibitemOpen
  \bibfield  {author} {\bibinfo {author} {\bibfnamefont {K.}~\bibnamefont
  {Schmidt-Hoberg}}, \bibinfo {author} {\bibfnamefont {F.}~\bibnamefont
  {Staub}}, \ and\ \bibinfo {author} {\bibfnamefont {M.~W.}\ \bibnamefont
  {Winkler}},\ }\href {\doibase 10.1016/j.physletb.2013.11.015} {\bibfield
  {journal} {\bibinfo  {journal} {Phys. Lett. B}\ }\textbf {\bibinfo {volume}
  {727}},\ \bibinfo {pages} {506} (\bibinfo {year} {2013}{\natexlab{b}})},\
  \Eprint {http://arxiv.org/abs/1310.6752} {arXiv:1310.6752 [hep-ph]}
  \BibitemShut {NoStop}%
\bibitem [{\citenamefont {Krnjaic}(2016{\natexlab{b}})}]{1512.04119}%
  \BibitemOpen
  \bibfield  {author} {\bibinfo {author} {\bibfnamefont {G.}~\bibnamefont
  {Krnjaic}},\ }\href {\doibase 10.1103/PhysRevD.94.073009} {\bibfield
  {journal} {\bibinfo  {journal} {Phys. Rev. D}\ }\textbf {\bibinfo {volume}
  {94}},\ \bibinfo {pages} {073009} (\bibinfo {year} {2016}{\natexlab{b}})},\
  \Eprint {http://arxiv.org/abs/1512.04119} {arXiv:1512.04119 [hep-ph]}
  \BibitemShut {NoStop}%
\bibitem [{\citenamefont {Fradette}\ and\ \citenamefont
  {Pospelov}(2017)}]{1706.01920}%
  \BibitemOpen
  \bibfield  {author} {\bibinfo {author} {\bibfnamefont {A.}~\bibnamefont
  {Fradette}}\ and\ \bibinfo {author} {\bibfnamefont {M.}~\bibnamefont
  {Pospelov}},\ }\href {\doibase 10.1103/PhysRevD.96.075033} {\bibfield
  {journal} {\bibinfo  {journal} {Phys. Rev. D}\ }\textbf {\bibinfo {volume}
  {96}},\ \bibinfo {pages} {075033} (\bibinfo {year} {2017})},\ \Eprint
  {http://arxiv.org/abs/1706.01920} {arXiv:1706.01920 [hep-ph]} \BibitemShut
  {NoStop}%
\bibitem [{\citenamefont {Jia}\ \emph {et~al.}(2022{\natexlab{a}})\citenamefont
  {Jia} \emph {et~al.}}]{2112.11852}%
  \BibitemOpen
  \bibfield  {author} {\bibinfo {author} {\bibfnamefont {S.}~\bibnamefont
  {Jia}} \emph {et~al.} (\bibinfo {collaboration} {Belle}),\ }\href {\doibase
  10.1103/PhysRevLett.128.081804} {\bibfield  {journal} {\bibinfo  {journal}
  {Phys. Rev. Lett.}\ }\textbf {\bibinfo {volume} {128}},\ \bibinfo {pages}
  {081804} (\bibinfo {year} {2022}{\natexlab{a}})},\ \Eprint
  {http://arxiv.org/abs/2112.11852} {arXiv:2112.11852 [hep-ex]} \BibitemShut
  {NoStop}%
\bibitem [{\citenamefont {Adachi}\ \emph
  {et~al.}(2023{\natexlab{b}})\citenamefont {Adachi} \emph
  {et~al.}}]{2306.02830}%
  \BibitemOpen
  \bibfield  {author} {\bibinfo {author} {\bibfnamefont {I.}~\bibnamefont
  {Adachi}} \emph {et~al.} (\bibinfo {collaboration} {Belle-II}),\ }\href
  {\doibase 10.1103/PhysRevD.108.L111104} {\bibfield  {journal} {\bibinfo
  {journal} {Phys. Rev. D}\ }\textbf {\bibinfo {volume} {108}},\ \bibinfo
  {pages} {L111104} (\bibinfo {year} {2023}{\natexlab{b}})},\ \Eprint
  {http://arxiv.org/abs/2306.02830} {arXiv:2306.02830 [hep-ex]} \BibitemShut
  {NoStop}%
\bibitem [{\citenamefont {Jia}\ \emph {et~al.}(2022{\natexlab{b}})\citenamefont
  {Jia} \emph {et~al.}}]{Belle:2021rcl}%
  \BibitemOpen
  \bibfield  {author} {\bibinfo {author} {\bibfnamefont {S.}~\bibnamefont
  {Jia}} \emph {et~al.} (\bibinfo {collaboration} {Belle}),\ }\href {\doibase
  10.1103/PhysRevLett.128.081804} {\bibfield  {journal} {\bibinfo  {journal}
  {Phys. Rev. Lett.}\ }\textbf {\bibinfo {volume} {128}},\ \bibinfo {pages}
  {081804} (\bibinfo {year} {2022}{\natexlab{b}})},\ \Eprint
  {http://arxiv.org/abs/2112.11852} {arXiv:2112.11852 [hep-ex]} \BibitemShut
  {NoStop}%
\bibitem [{\citenamefont {Ablikim}\ \emph {et~al.}(2022)\citenamefont {Ablikim}
  \emph {et~al.}}]{2109.12625}%
  \BibitemOpen
  \bibfield  {author} {\bibinfo {author} {\bibfnamefont {M.}~\bibnamefont
  {Ablikim}} \emph {et~al.} (\bibinfo {collaboration} {BESIII}),\ }\href
  {\doibase 10.1103/PhysRevD.105.012008} {\bibfield  {journal} {\bibinfo
  {journal} {Phys. Rev. D}\ }\textbf {\bibinfo {volume} {105}},\ \bibinfo
  {pages} {012008} (\bibinfo {year} {2022})},\ \Eprint
  {http://arxiv.org/abs/2109.12625} {arXiv:2109.12625 [hep-ex]} \BibitemShut
  {NoStop}%
\bibitem [{\citenamefont {Aaij}\ \emph {et~al.}(2015)\citenamefont {Aaij} \emph
  {et~al.}}]{1508.04094}%
  \BibitemOpen
  \bibfield  {author} {\bibinfo {author} {\bibfnamefont {R.}~\bibnamefont
  {Aaij}} \emph {et~al.} (\bibinfo {collaboration} {LHCb}),\ }\href {\doibase
  10.1103/PhysRevLett.115.161802} {\bibfield  {journal} {\bibinfo  {journal}
  {Phys. Rev. Lett.}\ }\textbf {\bibinfo {volume} {115}},\ \bibinfo {pages}
  {161802} (\bibinfo {year} {2015})},\ \Eprint
  {http://arxiv.org/abs/1508.04094} {arXiv:1508.04094 [hep-ex]} \BibitemShut
  {NoStop}%
\bibitem [{\citenamefont {Aaij}\ \emph {et~al.}(2017)\citenamefont {Aaij} \emph
  {et~al.}}]{1612.07818}%
  \BibitemOpen
  \bibfield  {author} {\bibinfo {author} {\bibfnamefont {R.}~\bibnamefont
  {Aaij}} \emph {et~al.} (\bibinfo {collaboration} {LHCb}),\ }\href {\doibase
  10.1103/PhysRevD.95.071101} {\bibfield  {journal} {\bibinfo  {journal} {Phys.
  Rev. D}\ }\textbf {\bibinfo {volume} {95}},\ \bibinfo {pages} {071101}
  (\bibinfo {year} {2017})},\ \Eprint {http://arxiv.org/abs/1612.07818}
  {arXiv:1612.07818 [hep-ex]} \BibitemShut {NoStop}%
\bibitem [{\citenamefont {Cortina~Gil}\ \emph
  {et~al.}(2021{\natexlab{b}})\citenamefont {Cortina~Gil} \emph
  {et~al.}}]{2011.11329}%
  \BibitemOpen
  \bibfield  {author} {\bibinfo {author} {\bibfnamefont {E.}~\bibnamefont
  {Cortina~Gil}} \emph {et~al.} (\bibinfo {collaboration} {NA62}),\ }\href
  {\doibase 10.1007/JHEP03(2021)058} {\bibfield  {journal} {\bibinfo  {journal}
  {JHEP}\ }\textbf {\bibinfo {volume} {03}},\ \bibinfo {pages} {058} (\bibinfo
  {year} {2021}{\natexlab{b}})},\ \Eprint {http://arxiv.org/abs/2011.11329}
  {arXiv:2011.11329 [hep-ex]} \BibitemShut {NoStop}%
\bibitem [{\citenamefont {Cortina~Gil}\ \emph
  {et~al.}(2021{\natexlab{c}})\citenamefont {Cortina~Gil} \emph
  {et~al.}}]{2010.07644}%
  \BibitemOpen
  \bibfield  {author} {\bibinfo {author} {\bibfnamefont {E.}~\bibnamefont
  {Cortina~Gil}} \emph {et~al.} (\bibinfo {collaboration} {NA62}),\ }\href
  {\doibase 10.1007/JHEP02(2021)201} {\bibfield  {journal} {\bibinfo  {journal}
  {JHEP}\ }\textbf {\bibinfo {volume} {02}},\ \bibinfo {pages} {201} (\bibinfo
  {year} {2021}{\natexlab{c}})},\ \Eprint {http://arxiv.org/abs/2010.07644}
  {arXiv:2010.07644 [hep-ex]} \BibitemShut {NoStop}%
\bibitem [{\citenamefont {Alavi-Harati}\ \emph {et~al.}(2004)\citenamefont
  {Alavi-Harati} \emph {et~al.}}]{hep-ex/0309072}%
  \BibitemOpen
  \bibfield  {author} {\bibinfo {author} {\bibfnamefont {A.}~\bibnamefont
  {Alavi-Harati}} \emph {et~al.} (\bibinfo {collaboration} {KTeV}),\ }\href
  {\doibase 10.1103/PhysRevLett.93.021805} {\bibfield  {journal} {\bibinfo
  {journal} {Phys. Rev. Lett.}\ }\textbf {\bibinfo {volume} {93}},\ \bibinfo
  {pages} {021805} (\bibinfo {year} {2004})},\ \Eprint
  {http://arxiv.org/abs/hep-ex/0309072} {arXiv:hep-ex/0309072} \BibitemShut
  {NoStop}%
\bibitem [{\citenamefont {Abouzaid}\ \emph {et~al.}(2008)\citenamefont
  {Abouzaid} \emph {et~al.}}]{0805.0031}%
  \BibitemOpen
  \bibfield  {author} {\bibinfo {author} {\bibfnamefont {E.}~\bibnamefont
  {Abouzaid}} \emph {et~al.} (\bibinfo {collaboration} {KTeV}),\ }\href
  {\doibase 10.1103/PhysRevD.77.112004} {\bibfield  {journal} {\bibinfo
  {journal} {Phys. Rev. D}\ }\textbf {\bibinfo {volume} {77}},\ \bibinfo
  {pages} {112004} (\bibinfo {year} {2008})},\ \Eprint
  {http://arxiv.org/abs/0805.0031} {arXiv:0805.0031 [hep-ex]} \BibitemShut
  {NoStop}%
\bibitem [{\citenamefont {Clarke}\ \emph {et~al.}(2014)\citenamefont {Clarke},
  \citenamefont {Foot},\ and\ \citenamefont {Volkas}}]{1310.8042}%
  \BibitemOpen
  \bibfield  {author} {\bibinfo {author} {\bibfnamefont {J.~D.}\ \bibnamefont
  {Clarke}}, \bibinfo {author} {\bibfnamefont {R.}~\bibnamefont {Foot}}, \ and\
  \bibinfo {author} {\bibfnamefont {R.~R.}\ \bibnamefont {Volkas}},\ }\href
  {\doibase 10.1007/JHEP02(2014)123} {\bibfield  {journal} {\bibinfo  {journal}
  {JHEP}\ }\textbf {\bibinfo {volume} {02}},\ \bibinfo {pages} {123} (\bibinfo
  {year} {2014})},\ \Eprint {http://arxiv.org/abs/1310.8042} {arXiv:1310.8042
  [hep-ph]} \BibitemShut {NoStop}%
\bibitem [{\citenamefont {Bezrukov}\ and\ \citenamefont
  {Gorbunov}(2010)}]{0912.0390}%
  \BibitemOpen
  \bibfield  {author} {\bibinfo {author} {\bibfnamefont {F.}~\bibnamefont
  {Bezrukov}}\ and\ \bibinfo {author} {\bibfnamefont {D.}~\bibnamefont
  {Gorbunov}},\ }\href {\doibase 10.1007/JHEP05(2010)010} {\bibfield  {journal}
  {\bibinfo  {journal} {JHEP}\ }\textbf {\bibinfo {volume} {05}},\ \bibinfo
  {pages} {010} (\bibinfo {year} {2010})},\ \Eprint
  {http://arxiv.org/abs/0912.0390} {arXiv:0912.0390 [hep-ph]} \BibitemShut
  {NoStop}%
\bibitem [{\citenamefont {Alekhin}\ \emph {et~al.}(2016)\citenamefont {Alekhin}
  \emph {et~al.}}]{1504.04855}%
  \BibitemOpen
  \bibfield  {author} {\bibinfo {author} {\bibfnamefont {S.}~\bibnamefont
  {Alekhin}} \emph {et~al.},\ }\href {\doibase 10.1088/0034-4885/79/12/124201}
  {\bibfield  {journal} {\bibinfo  {journal} {Rept. Prog. Phys.}\ }\textbf
  {\bibinfo {volume} {79}},\ \bibinfo {pages} {124201} (\bibinfo {year}
  {2016})},\ \Eprint {http://arxiv.org/abs/1504.04855} {arXiv:1504.04855
  [hep-ph]} \BibitemShut {NoStop}%
\bibitem [{\citenamefont {Gorbunov}\ \emph {et~al.}(2021)\citenamefont
  {Gorbunov}, \citenamefont {Krasnov},\ and\ \citenamefont
  {Suvorov}}]{2105.11102}%
  \BibitemOpen
  \bibfield  {author} {\bibinfo {author} {\bibfnamefont {D.}~\bibnamefont
  {Gorbunov}}, \bibinfo {author} {\bibfnamefont {I.}~\bibnamefont {Krasnov}}, \
  and\ \bibinfo {author} {\bibfnamefont {S.}~\bibnamefont {Suvorov}},\ }\href
  {\doibase 10.1016/j.physletb.2021.136524} {\bibfield  {journal} {\bibinfo
  {journal} {Phys. Lett. B}\ }\textbf {\bibinfo {volume} {820}},\ \bibinfo
  {pages} {136524} (\bibinfo {year} {2021})},\ \Eprint
  {http://arxiv.org/abs/2105.11102} {arXiv:2105.11102 [hep-ph]} \BibitemShut
  {NoStop}%
\bibitem [{\citenamefont {Elor}\ \emph {et~al.}(2023)\citenamefont {Elor},
  \citenamefont {McGehee},\ and\ \citenamefont {Pierce}}]{Elor:2021swj}%
  \BibitemOpen
  \bibfield  {author} {\bibinfo {author} {\bibfnamefont {G.}~\bibnamefont
  {Elor}}, \bibinfo {author} {\bibfnamefont {R.}~\bibnamefont {McGehee}}, \
  and\ \bibinfo {author} {\bibfnamefont {A.}~\bibnamefont {Pierce}},\ }\href
  {\doibase 10.1103/PhysRevLett.130.031803} {\bibfield  {journal} {\bibinfo
  {journal} {Phys. Rev. Lett.}\ }\textbf {\bibinfo {volume} {130}},\ \bibinfo
  {pages} {031803} (\bibinfo {year} {2023})},\ \Eprint
  {http://arxiv.org/abs/2112.03920} {arXiv:2112.03920 [hep-ph]} \BibitemShut
  {NoStop}%
\bibitem [{\citenamefont {Bhattiprolu}\ \emph {et~al.}(2023)\citenamefont
  {Bhattiprolu}, \citenamefont {Elor}, \citenamefont {McGehee},\ and\
  \citenamefont {Pierce}}]{Bhattiprolu:2022sdd}%
  \BibitemOpen
  \bibfield  {author} {\bibinfo {author} {\bibfnamefont {P.~N.}\ \bibnamefont
  {Bhattiprolu}}, \bibinfo {author} {\bibfnamefont {G.}~\bibnamefont {Elor}},
  \bibinfo {author} {\bibfnamefont {R.}~\bibnamefont {McGehee}}, \ and\
  \bibinfo {author} {\bibfnamefont {A.}~\bibnamefont {Pierce}},\ }\href
  {\doibase 10.1007/JHEP01(2023)128} {\bibfield  {journal} {\bibinfo  {journal}
  {JHEP}\ }\textbf {\bibinfo {volume} {01}},\ \bibinfo {pages} {128} (\bibinfo
  {year} {2023})},\ \Eprint {http://arxiv.org/abs/2210.15653} {arXiv:2210.15653
  [hep-ph]} \BibitemShut {NoStop}%
\bibitem [{\citenamefont {Parrott}\ \emph
  {et~al.}(2023{\natexlab{b}})\citenamefont {Parrott}, \citenamefont
  {Bouchard},\ and\ \citenamefont {Davies}}]{2207.12468}%
  \BibitemOpen
  \bibfield  {author} {\bibinfo {author} {\bibfnamefont {W.~G.}\ \bibnamefont
  {Parrott}}, \bibinfo {author} {\bibfnamefont {C.}~\bibnamefont {Bouchard}}, \
  and\ \bibinfo {author} {\bibfnamefont {C.~T.~H.}\ \bibnamefont {Davies}}
  (\bibinfo {collaboration} {(HPQCD collaboration)\textsection{}, HPQCD}),\
  }\href {\doibase 10.1103/PhysRevD.107.014510} {\bibfield  {journal} {\bibinfo
   {journal} {Phys. Rev. D}\ }\textbf {\bibinfo {volume} {107}},\ \bibinfo
  {pages} {014510} (\bibinfo {year} {2023}{\natexlab{b}})},\ \Eprint
  {http://arxiv.org/abs/2207.12468} {arXiv:2207.12468 [hep-lat]} \BibitemShut
  {NoStop}%
\bibitem [{\citenamefont {Bailey}\ \emph {et~al.}(2016)\citenamefont {Bailey}
  \emph {et~al.}}]{1509.06235}%
  \BibitemOpen
  \bibfield  {author} {\bibinfo {author} {\bibfnamefont {J.~A.}\ \bibnamefont
  {Bailey}} \emph {et~al.},\ }\href {\doibase 10.1103/PhysRevD.93.025026}
  {\bibfield  {journal} {\bibinfo  {journal} {Phys. Rev. D}\ }\textbf {\bibinfo
  {volume} {93}},\ \bibinfo {pages} {025026} (\bibinfo {year} {2016})},\
  \Eprint {http://arxiv.org/abs/1509.06235} {arXiv:1509.06235 [hep-lat]}
  \BibitemShut {NoStop}%
\bibitem [{\citenamefont {Gubernari}\ \emph {et~al.}(2019)\citenamefont
  {Gubernari}, \citenamefont {Kokulu},\ and\ \citenamefont {van
  Dyk}}]{Gubernari:2018wyi}%
  \BibitemOpen
  \bibfield  {author} {\bibinfo {author} {\bibfnamefont {N.}~\bibnamefont
  {Gubernari}}, \bibinfo {author} {\bibfnamefont {A.}~\bibnamefont {Kokulu}}, \
  and\ \bibinfo {author} {\bibfnamefont {D.}~\bibnamefont {van Dyk}},\ }\href
  {\doibase 10.1007/JHEP01(2019)150} {\bibfield  {journal} {\bibinfo  {journal}
  {JHEP}\ }\textbf {\bibinfo {volume} {01}},\ \bibinfo {pages} {150} (\bibinfo
  {year} {2019})},\ \Eprint {http://arxiv.org/abs/1811.00983} {arXiv:1811.00983
  [hep-ph]} \BibitemShut {NoStop}%
\bibitem [{\citenamefont {Aoki}\ \emph {et~al.}(2022)\citenamefont {Aoki} \emph
  {et~al.}}]{2111.09849}%
  \BibitemOpen
  \bibfield  {author} {\bibinfo {author} {\bibfnamefont {Y.}~\bibnamefont
  {Aoki}} \emph {et~al.} (\bibinfo {collaboration} {Flavour Lattice Averaging
  Group (FLAG)}),\ }\href {\doibase 10.1140/epjc/s10052-022-10536-1} {\bibfield
   {journal} {\bibinfo  {journal} {Eur. Phys. J. C}\ }\textbf {\bibinfo
  {volume} {82}},\ \bibinfo {pages} {869} (\bibinfo {year} {2022})},\ \Eprint
  {http://arxiv.org/abs/2111.09849} {arXiv:2111.09849 [hep-lat]} \BibitemShut
  {NoStop}%
\bibitem [{\citenamefont {Bharucha}\ \emph {et~al.}(2016)\citenamefont
  {Bharucha}, \citenamefont {Straub},\ and\ \citenamefont
  {Zwicky}}]{1503.05534}%
  \BibitemOpen
  \bibfield  {author} {\bibinfo {author} {\bibfnamefont {A.}~\bibnamefont
  {Bharucha}}, \bibinfo {author} {\bibfnamefont {D.~M.}\ \bibnamefont
  {Straub}}, \ and\ \bibinfo {author} {\bibfnamefont {R.}~\bibnamefont
  {Zwicky}},\ }\href {\doibase 10.1007/JHEP08(2016)098} {\bibfield  {journal}
  {\bibinfo  {journal} {JHEP}\ }\textbf {\bibinfo {volume} {08}},\ \bibinfo
  {pages} {098} (\bibinfo {year} {2016})},\ \Eprint
  {http://arxiv.org/abs/1503.05534} {arXiv:1503.05534 [hep-ph]} \BibitemShut
  {NoStop}%
\end{thebibliography}%
\end{document}